\documentclass[11pt]{article}

\usepackage{a4,amsmath,amssymb,here,epsf,bbold}

\newcommand\epsfig[4]{\begin{figure}[#2]
  \centerline{
    \epsfysize=#1
    \epsfbox{#3.eps}
    }
\vspace{0.2cm}
\caption{\em #4 \label{fig:#3}}  
\end{figure}
}

\newcommand\epsfigtwo[6]{\begin{figure}[#3]
  \centerline{
   a)  \epsfysize=#1
       \epsfbox{#4.eps} $ \quad \quad \quad $
   b)  \epsfysize=#2
       \epsfbox{#5.eps}
       }
\caption{{\em #6} \label{fig:#4}}  
\end{figure}
}

\newcommand\epsfigthreevarnew[8]{\begin{figure}[#4]
  \centerline{
   a)  \epsfysize=#1
       \epsfbox{#5.eps} $ \quad  $	
   b)  \epsfysize=#2
       \epsfbox{#6.eps} $ \quad  $	
   c)  \epsfysize=#3
       \epsfbox{#7.eps} $ \quad  $	}
\caption{{\em #8} \label{fig:#5}}  
\end{figure}
}

\newcommand\epsfigfour[8]{\begin{figure}[#3]
  \centerline{
   a)  \epsfysize=#1
       \epsfbox{#4.eps} $ \quad  $      
   b)  \epsfysize=#1
       \epsfbox{#5.eps} $ \quad  $      }
\centerline{
   c)  \epsfysize=#2
       \epsfbox{#6.eps} $ \quad  $      
   d)  \epsfysize=#2
       \epsfbox{#7.eps} $ \quad  $      }
\caption{\em #8 \label{fig:#4}}  
\end{figure}
}

\author{ Ma\'ira Aguiar$^+ $ \&  Nico Stollenwerk$^{\# }$\\ 
{\small Instituto Gulbenkian de Ci\^encias, }
\\{\small  Apartado 14, 
	2781-901 Oeiras, Portugal} 
\\{\small and}
\\{\small  
Faculdade de Ci\^encias and 
Centro de Matem\'atica }
\\{\small 
e Aplica\c{c}\~oes Fundamentais, Universidade de Lisboa, }
\\{\small 
Avenida Prof. Gama Pinto 2, 1649-003 Lisboa, Portugal}
\\{\footnotesize e-mails:  $^+$ {\tt maira@igc.gulbenkian.pt} 
and $^{\# }$ 
{\tt nks22@cam.ac.uk}  }
\\
}
\title{ 
         A new chaotic attractor in a basic multi-strain \\ epidemiological 
         model \\ with temporary cross-immunity
}
\date{{\small \today } }

\begin{document}
\maketitle
\vspace{0.0cm}

\begin{abstract}
\noindent
An epidemic multi-strain model 
with temporary cross-immunity shows
chaos, even in a previously unexpected parameter region. 
Especially dengue fever models with strong enhanced infectivity on 
secondary infection have previously shown deterministic chaos 
motivated by experimental
findings of antibody-dependent-enhancement (ADE).
Including temporary cross-immunity in such models, which is common
knowledge among field researchers in dengue, we find a deterministically
chaotic attractor in the more realistic parameter region of reduced
infectivity on secondary infection (''inverse ADE'' parameter region).
This is realistic for
dengue fever since on second infection people are more likely to be 
hospitalized, hence do not contribute to the force of infection as much as
people with first infection.

Our finding has wider implications beyond dengue in any multi-strain
epidemiological systems with altered 
infectivity uppon secondary infection, since we can relax the condition
of rather high infectivity on secondary infection previously required for
deterministic chaos.
For dengue the finding of wide ranges of chaotic attractors open new
ways to analysis of existing data sets.
\end{abstract}


\section{Introduction}


We observe deterministically chaotic attractors 
\cite{RuelleTakens,Ruelle89,Ott}
like the one
in Fig. \ref{fig:maximamaplight} for a multi-strain model with less infectivity
for secondary infection as for the first just by adding temporary
cross-immunity to previously existing dengue models.

Either researchers previously focussed in their models on higher infectivity
for secondary infection as for the first due to the hypothesized effect
of antibody-dependent-enhancement (ADE),
wich is confirmed in tissue experiments \cite{Halstead1982,Halstead2003},
to increase viral load on 
a secondary infection with a different strain than obtained in the
first infection \cite{FergusonN.et.al1999,Schwartz.et.al2005}.
Or they focussed on temporary cross-immunity versus ADE but again
limiting the effect of ADE to increase the contribution of secondary
cases to the force of infection \cite{HelenRohani2006}.

\epsfig{6cm}{htb}{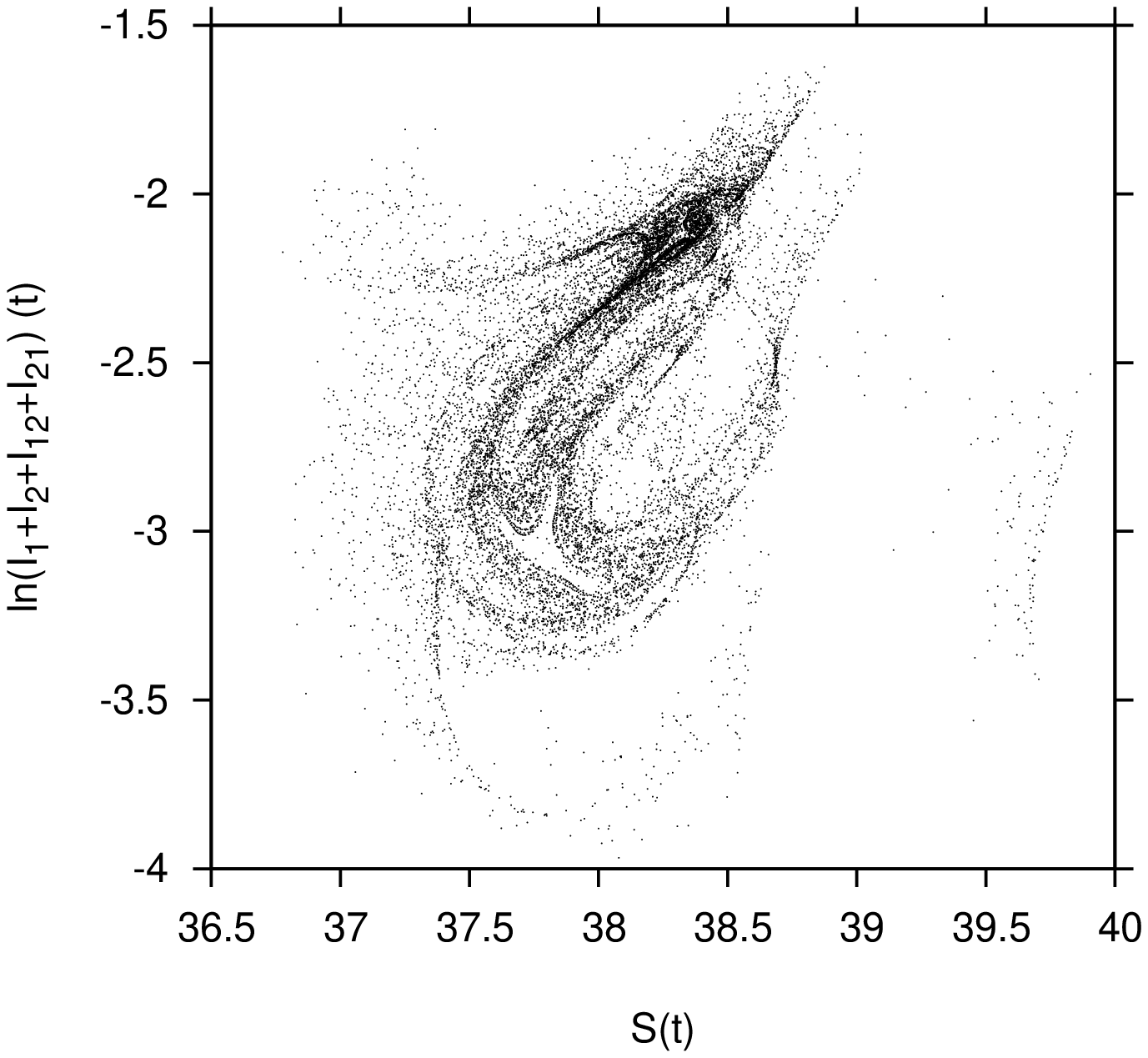}
	{Deterministically chaotic attractor obtained from a
          two-strain model with temporary cross-immunity
          in the ''inverse ADE'' parameter region of 
          contribution of secondary infection to the overall force
          of infection. A further description of this graphics 
          will be given in the next sections.
	}

Our model is a basic two-strain SIR-type model for the host population
only slightly refined as opposed to previously suggested models for
dengue fever. It is capturing the effective dynamics of the
human host population for the dengue virus, keeping effects of the mosquitoe
dynamics (dengue is transmitted by mosquitoes acting as transmission vectors
for the virus) or seasonality 
only in account by the effective parameters in the SIR-type model,
but not modelling these mechanisms explicitly. Instead we focus on the
multi-strain aspect and its effects on the host population.

This basic model structure allows to generalize our findings to other
multi-strain epidemiological systems, but is motivated by modelling
dengue fever epidemiology with its peculiar phenomenology.

Dengue is a mosquito-borne infection which in recent years 
has become a major international public health concern. 
Two fifths of the world's population is at risk from dengue. 
The World Health Organization (WHO) estimates there may be 
50 million cases of dengue infection worldwide every year. 
The disease is now endemic in over 100 countries throughout 
the Americas, South-East Asia, the western Pacific islands, 
Africa and the Eastern Mediterranean
\cite{WHO2002}.

Dengue fever is transmitted by the female domestic mosquito 
{\it Aedes aegypti}, although {\it Ae. albopictus} and {\it Ae. polynesiensis}
can also act as vector
\cite {Favier.et.al2005}.
Virus transmission in its simplest form involves the ingestion 
of viremic blood by mosquitoes and passage to a 
second susceptible human host.
The mosquito becomes infected when taking a blood meal 
from a viraemic person. After an extrinsic incubation 
period, the mosquito becomes infective 
and remains so during its entire life span
\cite {Rigau-Perez.et.al1998}. 

As the blood meal stimulates ovoposition, which undergoes 
at least one, often more, reproductive cycles there is 
an opportunity of vertical transmission to the eggs, 
passing the virus to the next generation of mosquitoes 
\cite {Rosen.et.al1983,Monath1994}.
In a dry state, eggs survive for very long periods 
and the virus can persist until the subsequent rainy season 
\cite {MortimerR}.

Dengue is caused by four antigenically distinct 
single-strand positive-polarity RNA viruses, 
designated dengue types $1$, $2$, $3$, and $4$, and belonging 
to the family Flaviviridae 
\cite{WHO2002}.
Infection by one serotype confers life-long immunity 
to only that serotype and temporary cross-immunity 
to other serotypes exists. It lasts from three to  nine months, 
when the antibody levels created during the response 
to that infection would be enough to protect against infection 
by a different but related serotype
\cite{SES,HelenRohani2006}.

Two forms of the disease exist: dengue fever (DF) 
or classic dengue, often benign, 
and dengue hemorrhagic fever (DHF), which may evolve 
towards a severe form known as dengue shock syndrome (DSS)
\cite {DerouichBoutayebTwizell2003}.
Without proper treatment DHF case fatality rates can exceed 20$ \% $
\cite{WHO2002}.
DF is characterized by headache, retro-orbital pain, 
myalgia, arthralgia, rash, leukopenia, and mild thrombocytopenia. 
The symptoms resolve after 2--7 days. DHF is a 
potentially deadly complication that is characterized 
by high fever and haemorrhagic phenomenae. DHF develops rapidly, 
usually over a period of hours, and resolves within 1--2 days 
in patients who receive appropriate fluid resuscitation. 
Otherwise, it can quickly progress to shock 
\cite{Huan-Yao.et.al2001}.

There are indeed pre-existing antibodies to previous dengue virus 
that cannot neutralize but rather enhance infection in vitro, a process described as antibody-dependent enhancement.
Epidemiological studies support the association of 
DHF and DSS with secondary dengue infection.
Halstead \cite{WHO2002,Halstead1982,Halstead2003} 
found that DHF and DSS were 15-80 times more likely 
in secondary then in primary infections and were 
positively associated with pre-existing dengue-virus-specific antibodies. 
However, there is no animal model of DHF and DSS, 
and the the causal relationship between ADE and 
severe disease remains unverified 
\cite{Huan-Yao.et.al2001}.

There is no specific treatment for dengue. 
A vaccine against dengue is not yet available, 
since it would have to simulate a protective immune response 
to all four serotypes
\cite{StephensonJR2005}.

Mathematical models describing the transmission of dengue viruses 
appeared in the literature as early as 1970 
\cite{FischerHalstead1970}.
More recently modelling attention has focussed on including ADE
as due to higher viral load of hosts on secondary infection
than on the first, hence a higher contribution to the force of
infection of each strain,
reporting deterministically chaotic attractors \cite{FergusonN.et.al1999}
and chaos de-synchronization \cite{Schwartz.et.al2005} to
explain the co-existence of the known four dengue viral strains.
Temporary cross-immunity against all strains after a first infection
has been included in mathematical models as well \cite{HelenRohani2006},
but to our knowledge, no systematic investigation of the attractor
structures of simple two-strain models with dengue-realistic
temporary cross-immunity and decreased contribution of secondary
infection to the force of infection, due to severity of infection with a second strain
and eventual hospitalization, has been 
performed so far. 

Our finding of chaotic attractors in this
parameter region thus indicate that deterministic chaos is
wider present than previously expected in multi-strain models.
For such scenarios new tools of non-linear data analysis like
Takens' embedding are available \cite{Packardetal80,Takens81}, 
and allow to obtain topological information (fixed points,
periodic orbits and the nature of chaotic attractors)
about the whole multi-strain epidemiological system
from time series of overall infecteds only, not needing 
any single strain data sets.

In the next section we present the two-strain model and give the
deteministic mean field version of it.
We then analyze the deterministic ODE-model in various parameter regions in
its attractor structure.



\section{The basic epidemic model }


Multi-strain dynamics are modelled with SIR-type models,
dividing the host population into susceptible, infected and recovered
individuals.

In the simple SIR epidemics without strain structure of the
pathogens we have the following reaction scheme
for the possible transitions from one to another state
\begin{eqnarray}
	S+I	& \stackrel{\beta  }{\longrightarrow }  &  	I+I
	\nonumber
\\
	I 	& \stackrel{\gamma }{\longrightarrow }  &  		R
	\nonumber
\\
	R	& \stackrel{\alpha }{\longrightarrow }  &  		S
	\nonumber
\\
	S,I,R	& \stackrel{\mu }{\longrightarrow }  &  		S
	\nonumber
\end{eqnarray}
for a host population of $N $ individuals, 
with contact and infection rate $\beta $, recovery rate $\gamma $
and temporary immunity rate $\alpha $.
Demography is denoted in the last reaction as exits from all
classes $S $, $I $ and $R $ with rate $\mu $ to the new born
susceptibles.

To include population noise stochastic models are investigated.
For state vectors $\underline n$, here  for the SIR-model 
$\underline n= (S,I,R)$, the master equation \cite{vanKampen} reads
\begin{equation}
	{\frac{d p(\underline n )}{d t }}
 		= \sum_{\underline{\tilde n} \neq \underline n } 
	\;w_{\underline n,\underline{\tilde n}} \;\; p(\underline{\tilde n}) -
		\sum_{\underline{\tilde n} \neq \underline n }
		\;w_{\underline{\tilde n}, \underline n} \;\; p(\underline n)
	\label{masterequ}
\end{equation}
with transition rates $w_{\underline{\tilde n}, \underline n} $ given by
the following expressions
\begin{eqnarray}
w_{(S+1,I,R-1), (S,I,R)} & = &  \alpha \cdot R
	\nonumber
\\
w_{(S-1,I+1,R), (S,I,R)} & = &  \beta  \cdot \frac{I}{N} \; S
\\
w_{(S,I-1,R+1), (S,I,R)} & = &  \gamma \cdot I
	\nonumber
\end{eqnarray}
from which the rates $w_{\underline n,\underline{\tilde n}} $ 
follow immediately as
\begin{eqnarray}
w_{(S,I,R), (S-1,I,R+1)} & = &  \alpha \cdot (R+1)
	\nonumber
\\
w_{(S,I,R), (S+1,I-1,R)} & = &  \beta  \cdot \frac{I-1}{N} \; (S+1)
\\
w_{(S,I,R), (S,I+1,R-1)} & = &  \gamma \cdot (I+1)
	\nonumber
	\quad.
\end{eqnarray}
In addition we have the transitions for the demography with rate $\mu $,
e.g. for the transition from recovered to susceptibles 
$w_{(S+1,I,R-1), (S,I,R)} =  \mu \cdot R $.
This formulation defines the stochastic process completely to capture
demographic noise
and will be the basis 
for the multi-strain model to be described in the following.

The above mentioned deterministic ODE model describes in mean field
approximation the dynamics of the mean values, e.g.
\begin{equation}
 \langle I  \rangle		
        :=  
        \sum_{I=0}^{N} \; I \; p(S,I,R)
        \quad.
        \label{definitionmeanvalue}
\end{equation}
The dynamics for the mean value is then given by inserting the
master equation
\begin{equation}
 \frac{d}{dt} \langle I  \rangle		
        =         \frac{\beta }{N} \cdot  
                           \langle   I\cdot S \rangle 
                           - (\gamma +\mu ) \cdot  \langle I  \rangle
        \approx 
                 \frac{\beta }{N} \cdot  
                           \langle   I \rangle \cdot \langle  S \rangle 
                           - (\gamma +\mu ) \cdot  \langle I  \rangle
        \label{dynamicsmeanvalue}
\end{equation}
with the mean field approximation 
$\langle   I\cdot S \rangle \approx 
\langle   I \rangle \cdot \langle  S \rangle $.
Carried out for all mean values $\langle  S \rangle $, $\langle   I\rangle  $
and  $\langle  R \rangle  $ gives a  closed ODE system for the SIR dynamics,
which is a deterministic system in the sense that initial values
determin the time course of the system for all times.
The mean field ODE system reads, now omitting the brackets for
mean values,
\begin{eqnarray}
    \frac{dS}{dt}& = &  \alpha R     -\frac{\beta }{N} \cdot  
                             I\cdot S  + \mu (N-S)\nonumber \\
    \frac{dI}{dt}& = &       \frac{\beta }{N} \cdot  
                             I\cdot S- \gamma  I  - \mu I \\
    \frac{dR}{dt}& = &     \gamma  I  -\alpha R  -\mu R \nonumber 
\end{eqnarray}
and has only fixed points as possible stationary solutions, i.e.
attractors. Transients under certain parameter values oscillate
into the fixed point, hence can be already more complex than the
final attractor. Stochastic versions of such models with only
fixed points possible as attractors but oscillating transients are
reported to also show stabilization of the oscillations due to
population noise \cite{AlonsoMcKanePascual,McKaneNewman}. 
So the deterministic mean field ODE models with their attractors give rather
the minimal complexity caused by the model structure.


\subsection{Basic two-strain model }


For two strains, $1$ and $2$, we have the following SIR-type model, now labelling the SIR classes for the hosts that have seen the individual strains.
Susceptibles to both strains (S) get infected with strain $1$ ($I_1$) or strain $2$ ($I_2$), with force of infection $\beta_1$ and $\beta_2$ respectively. They recover from infection with strain $1$ (becoming $R_1$) or from strain $2$ (becoming $R_2$), with recovery rate $\gamma$. 
In this recovered class, people have full immunity against the strain that they were exposed to and infected, and also, temporary immunity against the other strain (called period of temporary cross-immunity). After this, with rate $\alpha$, they enter again in the susceptible classes ($S_1$ respectively $S_2$), where the index represents the first infective strain. Now, $S_1$ can be reinfected with strain $2$ ($I_{12}$) and $S_2$ can be reinfected with strain $1$ ($I_{12}$), with  infection rates $\phi_2\beta_2$ and $\phi_1\beta_1$. The parameter $\phi$ in our model acts as "inverse ADE parameter", decreasing the infectivity of secondary infection, where people are more likely to be hospitalized because of the severity of the disease (DHF), do not contributing to the force of infection as much as people with first infection do. Finally, $I_{12}$ and $I_{12}$ go to the recovered class (R), immune against all strains. We include demography of the host population by denoting the birth and death rate  by $\mu $, assuming constant 
population size $N$.

In the following reaction scheme we describe the transitions for first infection with strain $1$ and secondary infection with strain $2$
\begin{eqnarray}
	S+I_1	& \stackrel{\beta _1 }{\longrightarrow }  &  	I_1+I_1
	\nonumber
\\
	S+I_{21} & \stackrel{\phi _1\beta _1 }{\longrightarrow }  &  I_1+I_{21}
	\nonumber
\\
	I_1 	& \stackrel{\gamma }{\longrightarrow }  &  	R_1
	\nonumber
\\
	R_1	& \stackrel{\alpha }{\longrightarrow }  &  	S_1
	\nonumber
\\
	S_1+I_2	& \stackrel{\beta _2 }{\longrightarrow }  &  	I_{12}+I_2
	\nonumber
\\
      S_1+I_{12} & \stackrel{\phi _2\beta _2 }{\longrightarrow }  & I_{12}+I_{12}
	\nonumber
\\
	I_{12} 	& \stackrel{\gamma }{\longrightarrow }  &  	R
	\nonumber
\end{eqnarray}
and for the reverse process, where the first infection is caused by strain $2$ and the secondary infection is caused by strain $1$.
\begin{eqnarray}
	S+I_2	& \stackrel{\beta _2 }{\longrightarrow }  &  	I_2+I_2
	\nonumber
\\
	S+I_{12} & \stackrel{\phi _2\beta _2 }{\longrightarrow }  &  I_2+I_{12}
	\nonumber
\\
	I_2 	& \stackrel{\gamma }{\longrightarrow }  &  	R_2
	\nonumber
\\
	R_2	& \stackrel{\alpha }{\longrightarrow }  &  	S_2
	\nonumber
\\
	S_2+I_1	& \stackrel{\beta _1 }{\longrightarrow }  &  	I_{21}+I_1
	\nonumber
\\
      S_2+I_{21} & \stackrel{\phi _1\beta _1 }{\longrightarrow }  & I_{21}+I_{21}
	\nonumber
\\
	I_{21} 	& \stackrel{\gamma }{\longrightarrow }  &  	R
	\nonumber
        \quad.
\end{eqnarray}
The demographic transitions are
\begin{eqnarray}
S,I_1,I_2,R_1,R_2,S_1,S_2,I_{12},I_{21},R	
& \stackrel{\mu }{\longrightarrow }  &  S
	\nonumber
\end{eqnarray}
defining the system of two strains completely. 

The stochastic version is now in complete analogy to the previous SIR model, and we can calculate the mean field equations (or simply read it off from the reaction scheme directly).

The mean field ODE system for the two strain epidemiological system is 
\begin{eqnarray}
    \frac{dS}{dt}& = &   - \frac{\beta_1}{N} S (I_1+ \phi_1 I_{21}) - \frac{\beta_2}{N} S (I_2 + \phi_2 I_{12}) + \mu (N-S) \nonumber
\\\nonumber
\\\nonumber
	\frac{dI_1}{dt}& = &   \frac{\beta_1}{N}S (I_1+ \phi_1 I_{21}) - (\gamma + \mu)I_1 \nonumber
\\\nonumber
\\\nonumber
	\frac{dI_2}{dt}& = &   \frac{\beta_2}{N}S (I_2+ \phi_2 I_{12}) - (\gamma + \mu)I_2 \nonumber
\\\nonumber
\\\nonumber
	\frac{dR_1}{dt}& = &   \gamma I_1 - (\alpha +\mu) R_1  \nonumber
\\\nonumber
\\
	\frac{dR_2}{dt}& = &   \gamma I_2 - (\alpha + \mu ) R_2 
        \label{ODE2strain}
\\ \nonumber
\\\nonumber
	\frac{dS_1}{dt}& = &  - \frac{\beta_2}{N} S_1 (I_2 + \phi_2 I_{12}) + \alpha R_1 - \mu S_1 \nonumber
\\\nonumber
\\\nonumber
	\frac{dS_2}{dt}& = &   - \frac{\beta_1}{N} S_2 (I_1 + \phi_1 I_{21}) + \alpha R_2 - \mu S_2\nonumber
\\\nonumber
\\\nonumber
	\frac{dI_{12}}{dt}& = &   \frac{\beta_2}{N}S_1 (I_2+ \phi_2 I_{12}) - (\gamma + \mu)I_{1,2}\nonumber
\\\nonumber
\\\nonumber
	\frac{dI_{21}}{dt}& = &   \frac{\beta_1}{N}S_2 (I_1+ \phi_1 I_{21}) - (\gamma + \mu)I_{21}\nonumber
\\\nonumber
\\\nonumber
	\frac{dR}{dt}& = &   \gamma (I_{12} +  I_{21}) - \mu R \nonumber
\end{eqnarray}
We consider $\phi _1 = \phi_2 = \phi $, $\beta _1 =\beta _2 = \beta $
for the moment, hence no epidemiological assymetry between the strains.

To take biological information from experiences in dengue
into account we fixe the transition rates of the model
as far as is known, and only will vary the unknown parameters
$\phi $ and eventually $\alpha $. Future work will have to also
adjust the other parameters better to describe actual data
of dengue cases.
The parameter values are, if not otherwise explicitly stated,
$\mu = 1/65 years $, $\gamma =52 y^{-1} $, $\beta = 2\cdot \gamma $
and $\alpha = 2 y^{-1} $. For the chaotic attractors we take the exemplaric values 
$\phi =0.7 $ and $\phi =2.7 $. We will also vary $\phi $ continuously
to obtain bifurcation diagrams \cite{Ruelle89,Ott}.



\section{Analysis of the multi-strain model with temporary
cross-immunity}


We will first look at time series simulations of the present model, 
Eq. system (\ref{ODE2strain}), and from there  we performed a detailed analysis of chaos, investigating state
space plots for various values for $\phi $, observing a rich structure
of attractors from fixed points to bifurcating limit cycles and
chaotic attractors. Finally, we will show whole bifurcation diagrams
for a systematic analysis for the entire parameter region of
$\phi  $ for various values of the temporary cross-immunity $\alpha $.


\subsection{Time series for {\boldmath $\phi > 1 $}}


In order to classify the dynamic pattern of the model for various
parameters we discard long transients which would carry some information
of the initial conditions. So we neglect the first 2000 years in the
following simulations.
We first simulate time series for $\phi>1$, 
see Fig. \ref{fig:timeseriesSphi27} and Fig. \ref{fig:timeseriesIphi27}.

\epsfigthreevarnew{3cm}{3cm}{3cm}{htb}{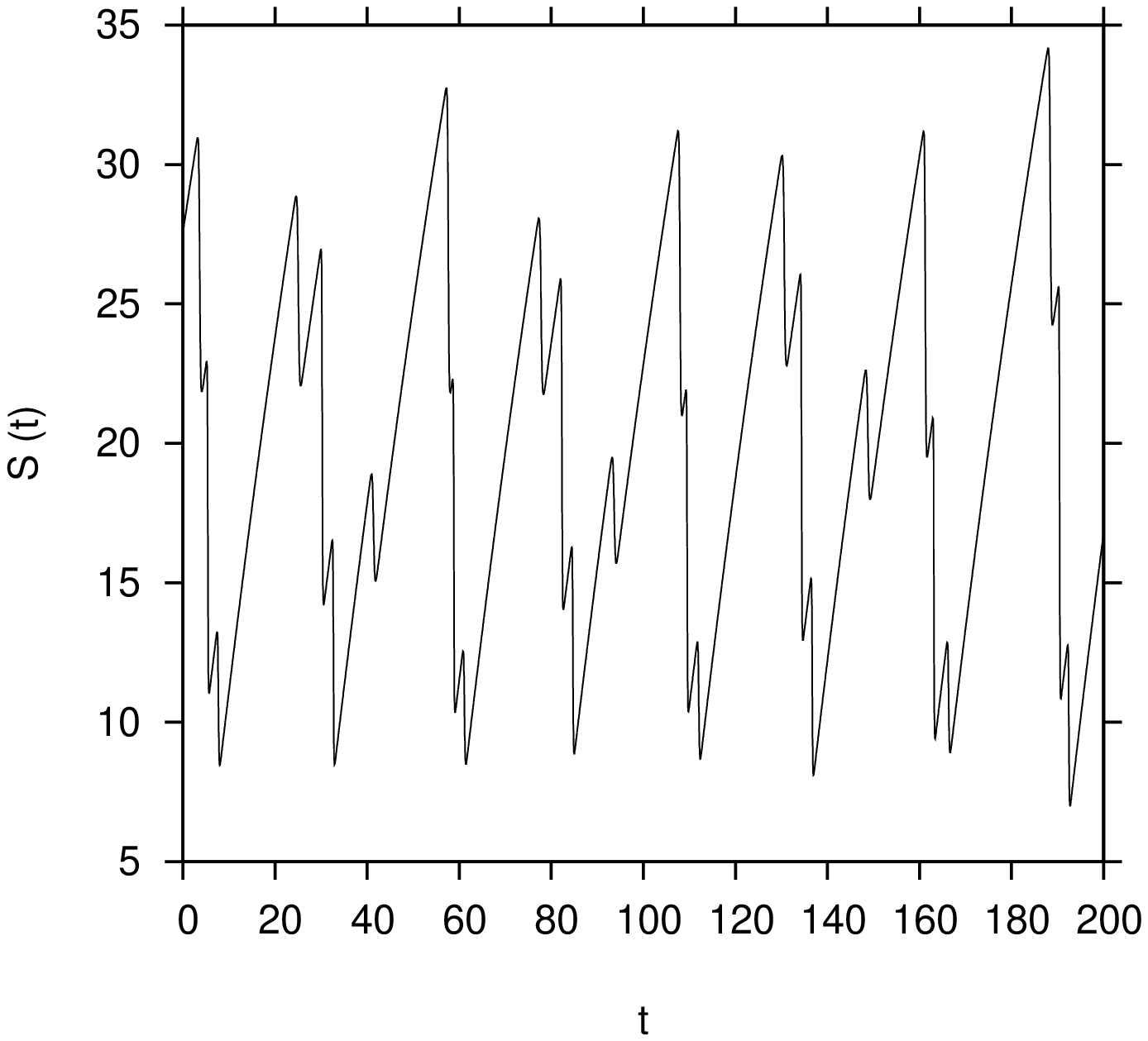}{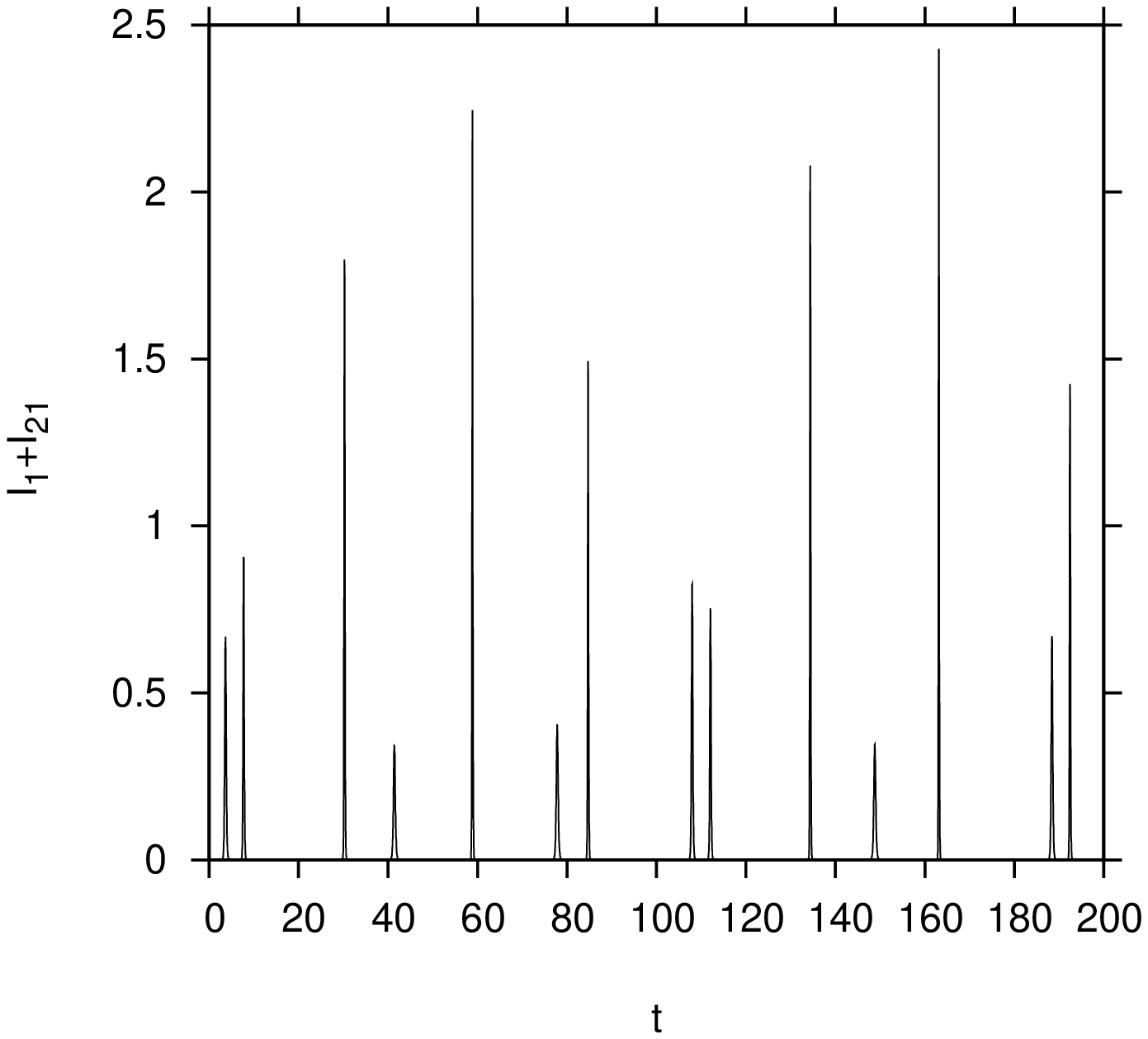}{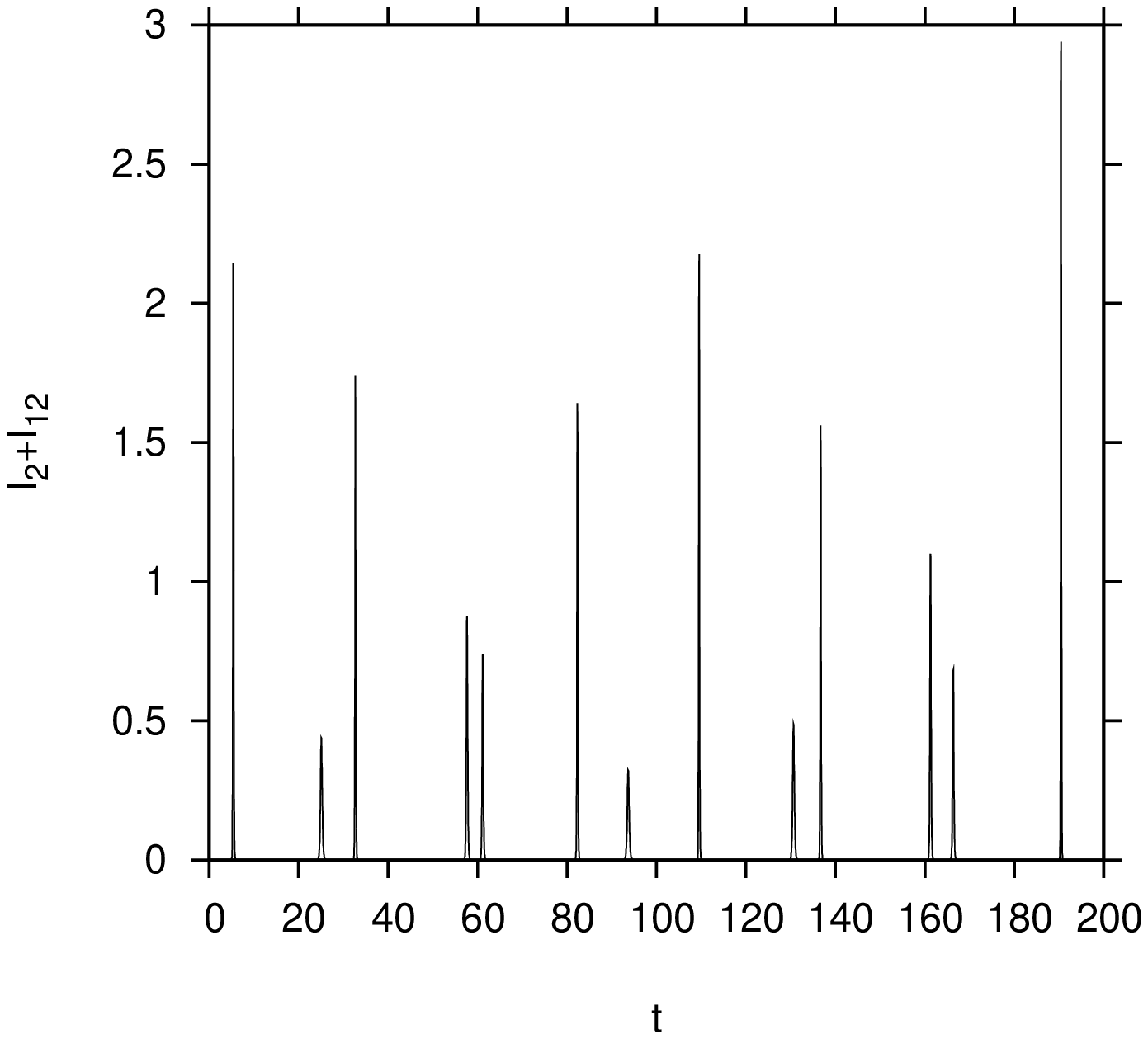}
	{For $\phi=2.7 $, time series simulations of a) the suceptibles $S $,
          b) the at the moment infected with strain 1 ($I_1+I_{21} $),
          c) the at the moment infected with strain 2 ($I_2+I_{12} $).
          Temporary cross-immunity is assumed to be
          $\alpha=2 $.
          The absolute numbers on the y-achses indicate percentage
          ($N=100 $).
	}

\epsfigtwo{4cm}{4cm}{htb}{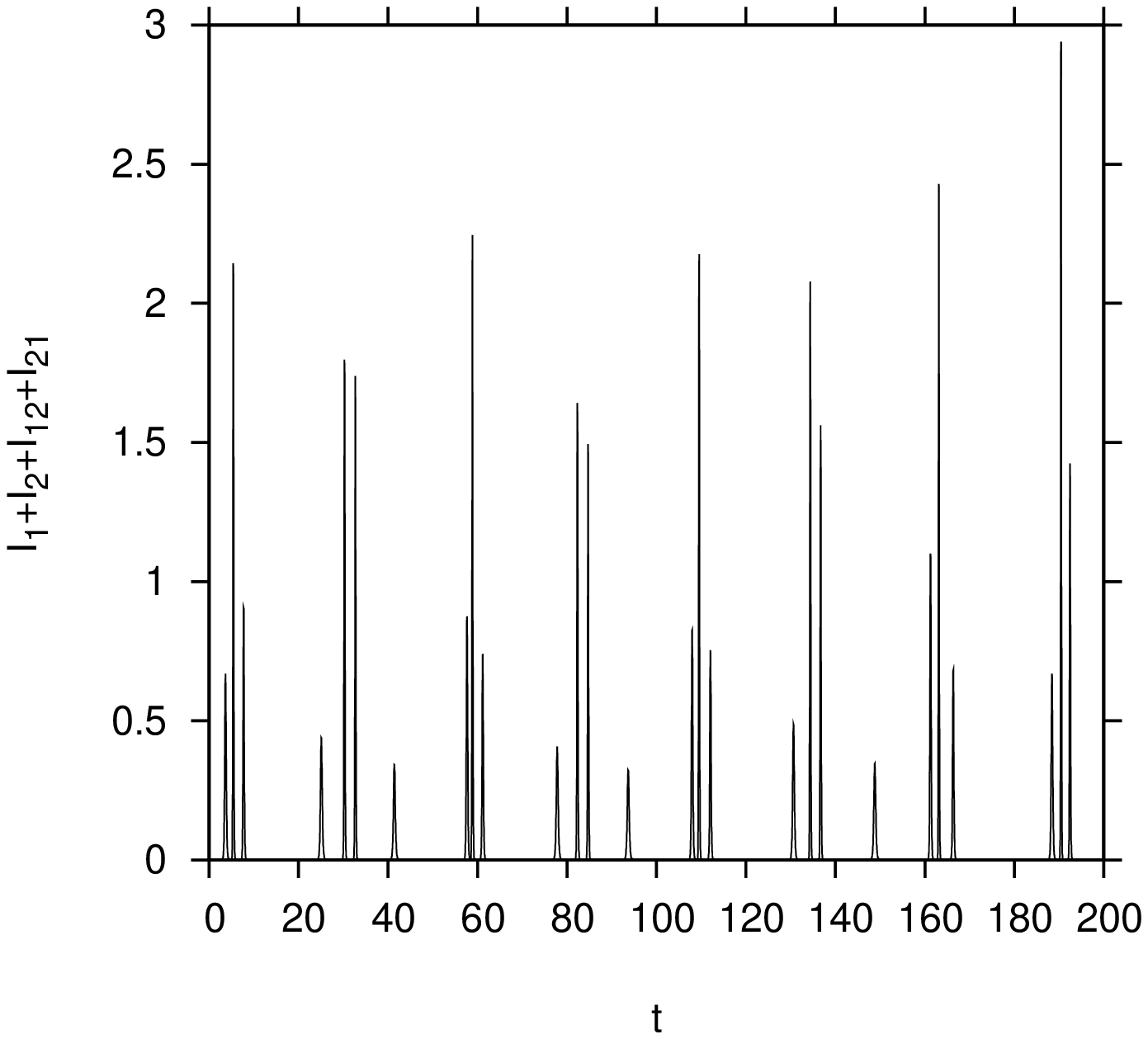}{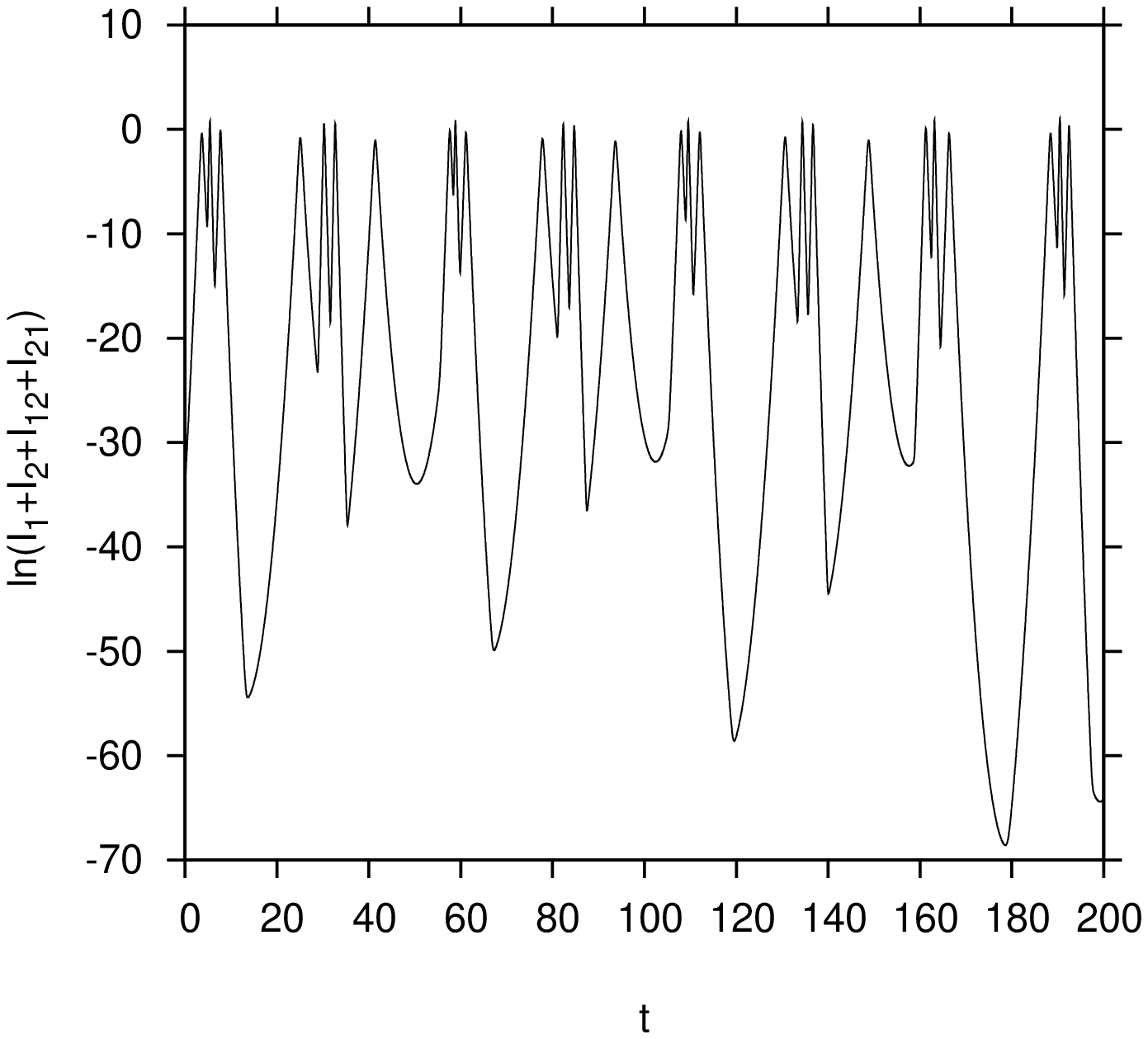}
	{Using the same parameter values above, In a) we plot the time series of the total number of infected
          $I:=I_1+I_2+I_{12}+I_{21} $, and in
          b) time series of the logarithm of the
          overall infected (ln$(I)$), for the same time
          interval.
          Very deep troughs are observed for these parameter values.
	}

However, also the transients reflect the dynamic behaviour of the system
under the present parameter values. Spiraling into a fixed point often
indicates a nearby periodic orbit. Irregular transients indicate
chaotic behaviour in neighbouring parameter regions etc. Hence, even
if dengue fever would be evolutionarily younger 
than the mentioned 2000 years of
transient, the observed pattern would still give information about the
dynamics of the system.
Also the stochastic version of this deterministic mean field model would 
be rather more complex than the attractor classification 
and not simpler \cite{AlonsoMcKanePascual,McKaneNewman},
of course besides extinction due to population fluctuations for
eventually very low numbers of infected.

Rather low troughs are observed for the total number of infected leading to
unrealistically low numbers of infected on average in low epidemic times.
The present deterministic model can be interpreted as a mean field
model for stochastic models capturing the effect of population noise
e.g. in terms of master equations 
as mentioned before
\cite{vanKampen}.
The ODE solutions are then the mean values of such stochastic models.
Population fluctuations would in the present case drive always the system
to extinction.



\subsection{Time series for {\boldmath $\phi < 1 $}}


Investigating time series for $\phi < 1 $, as would be realistic for dengue
fever due to more severe disease upon reinfection, hence larger
chance of being hospitalized, also indicates more complicated dynamic
behaviour than just simple fixed point or limit cycles. Discarding the first $2000$ years, see Fig. \ref{fig:timeseriesSphi07} 
and Fig. \ref{fig:timeseriesIphi07}.

\epsfigthreevarnew{3cm}{3cm}{3cm}{htb}{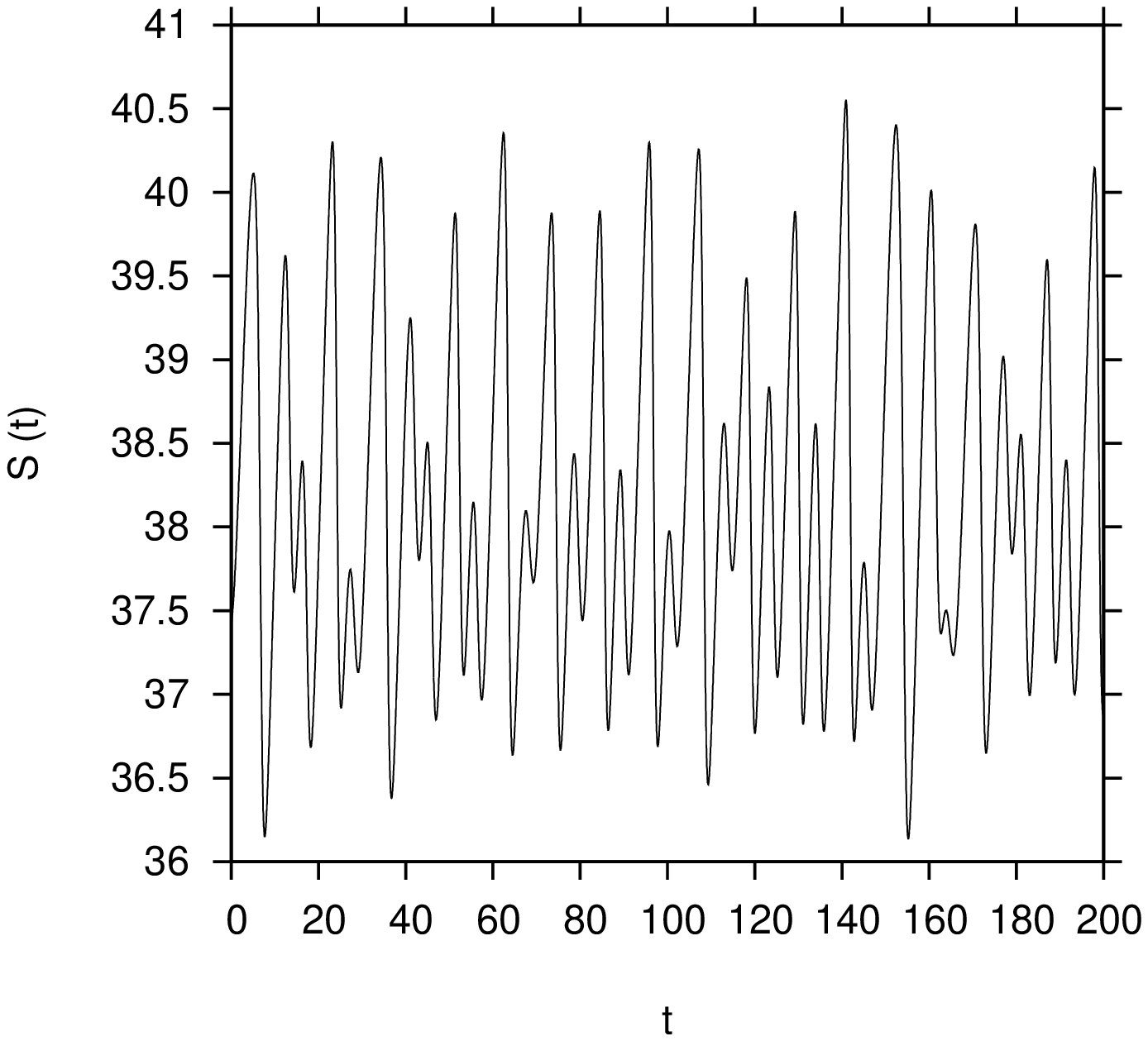}{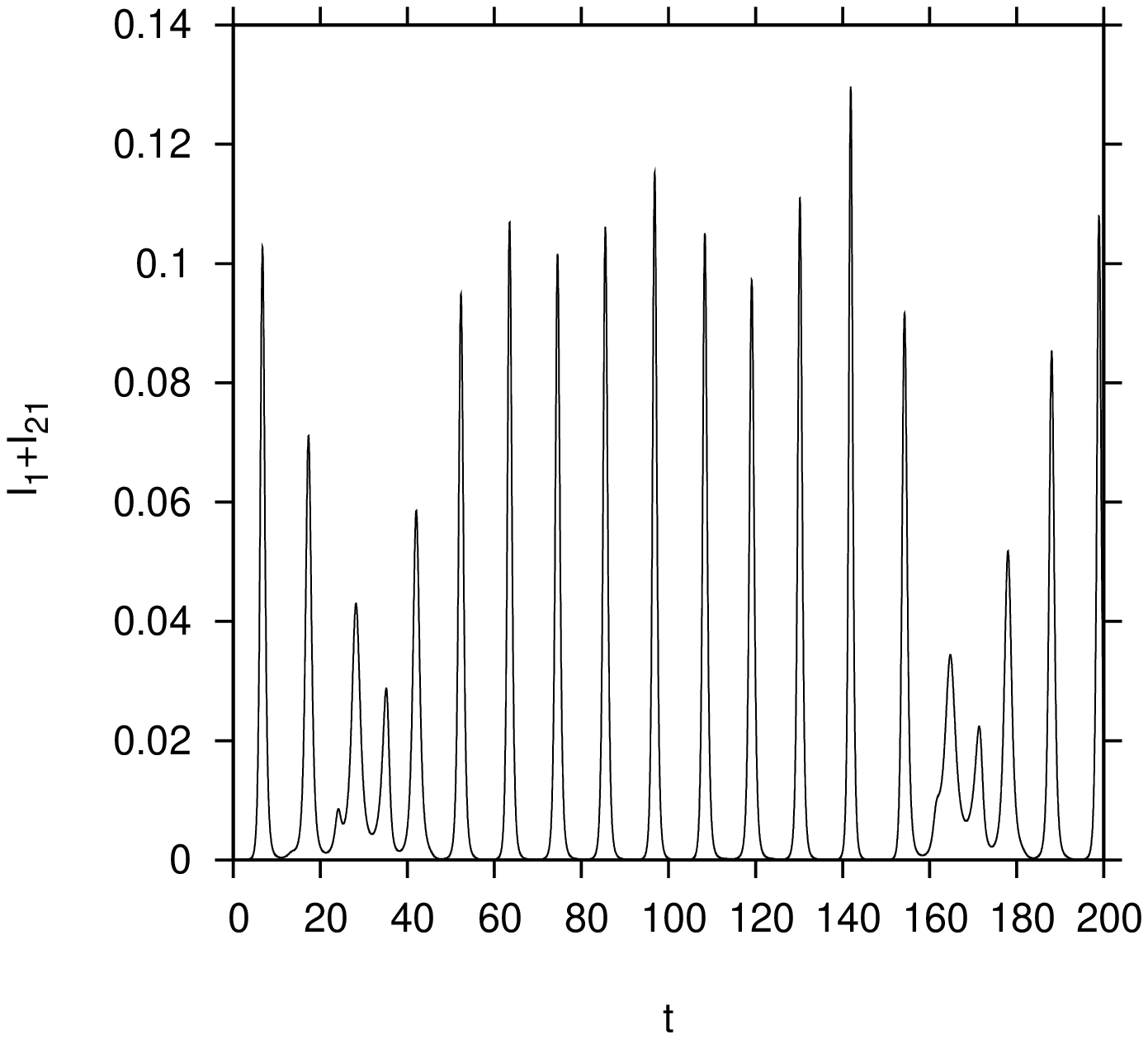}{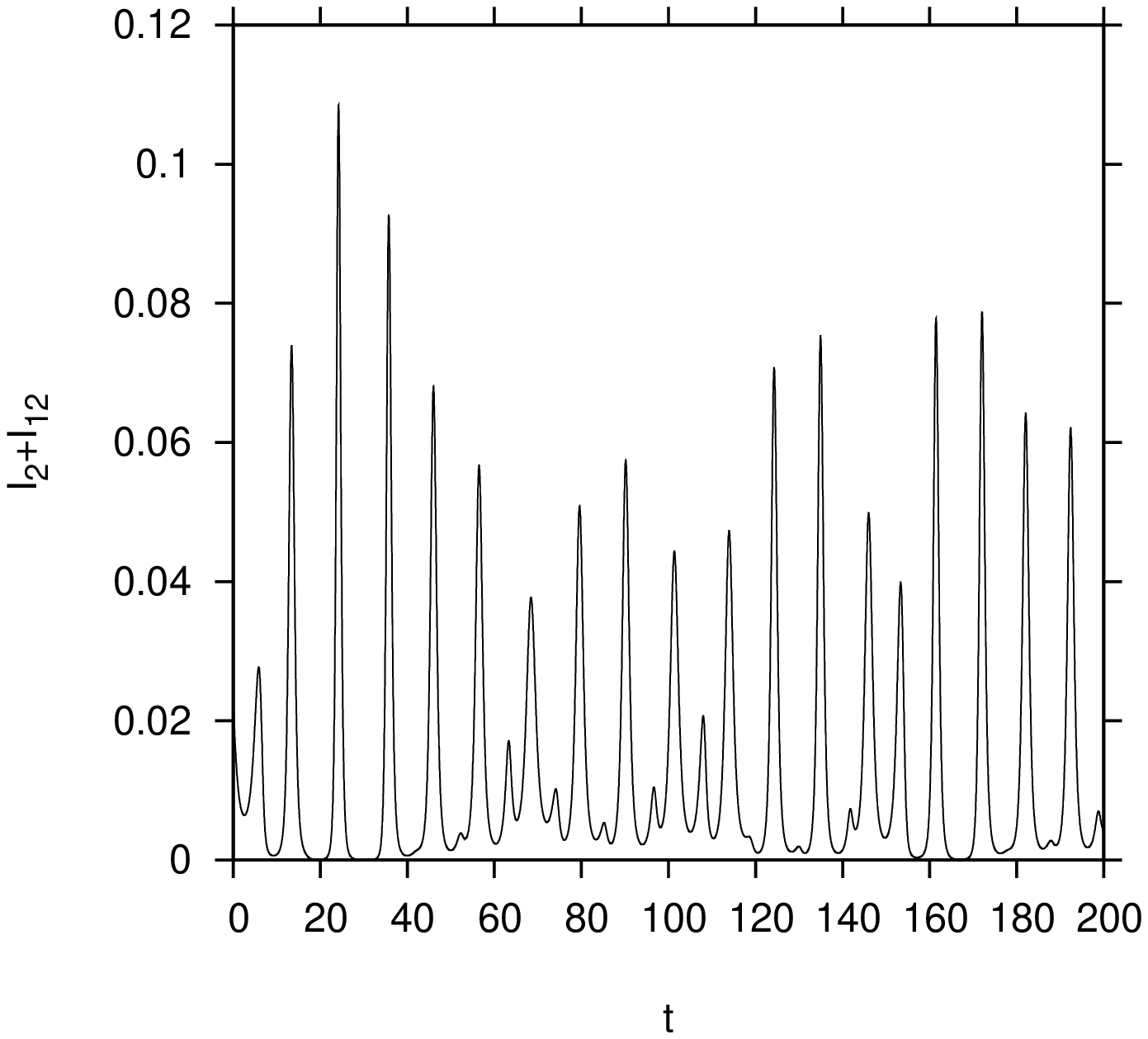}
	{For $\alpha=2$ and $\phi=0.7 $, time series simulations of a) the suceptibles $S $,
          b) the at the moment infected with strain 1 ($I_1+I_{21} $), and 
          c) the at the moment infected with strain 2 ($I_2+I_{12} $).
	}

\epsfigtwo{4cm}{4cm}{htb}{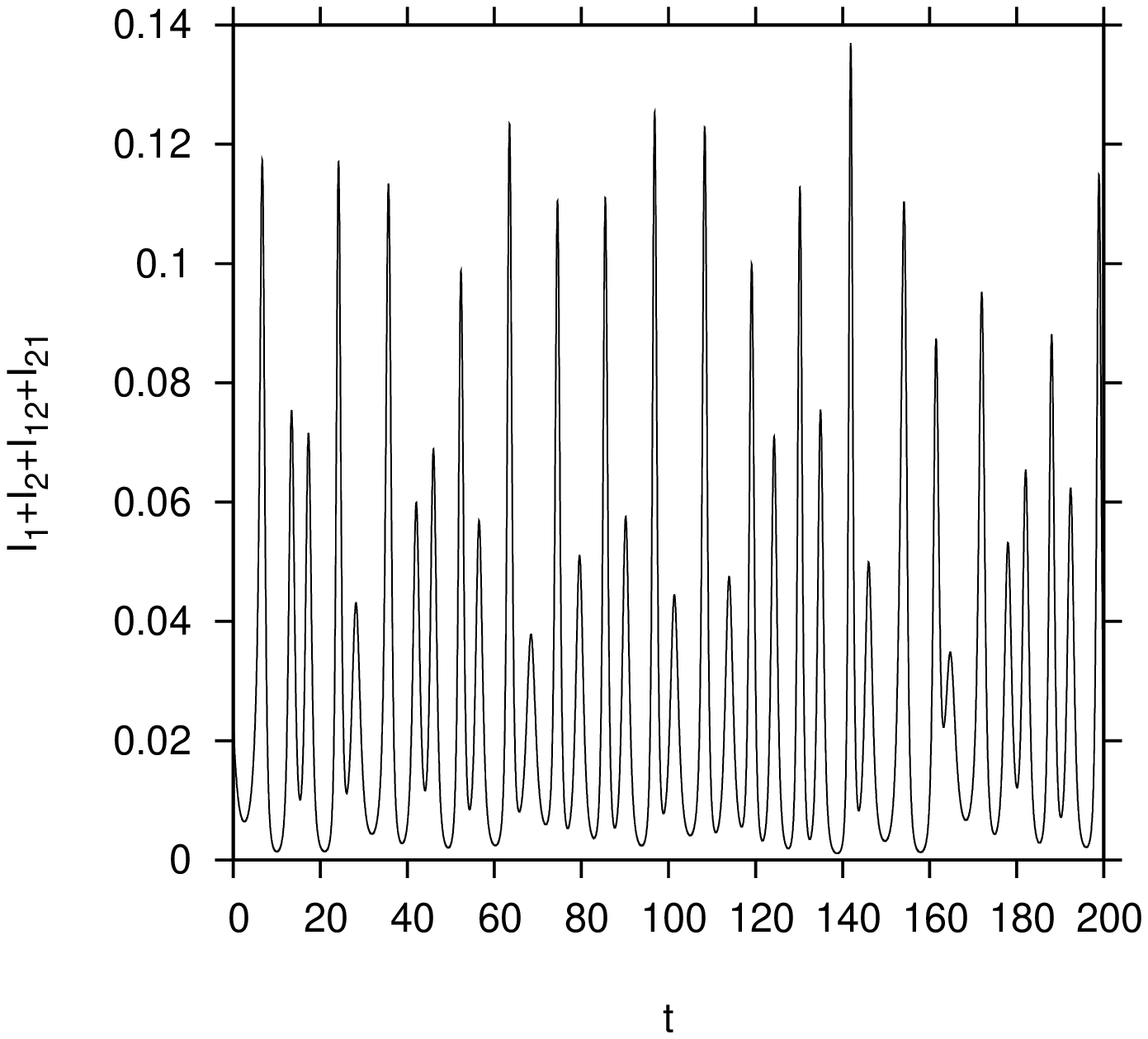}{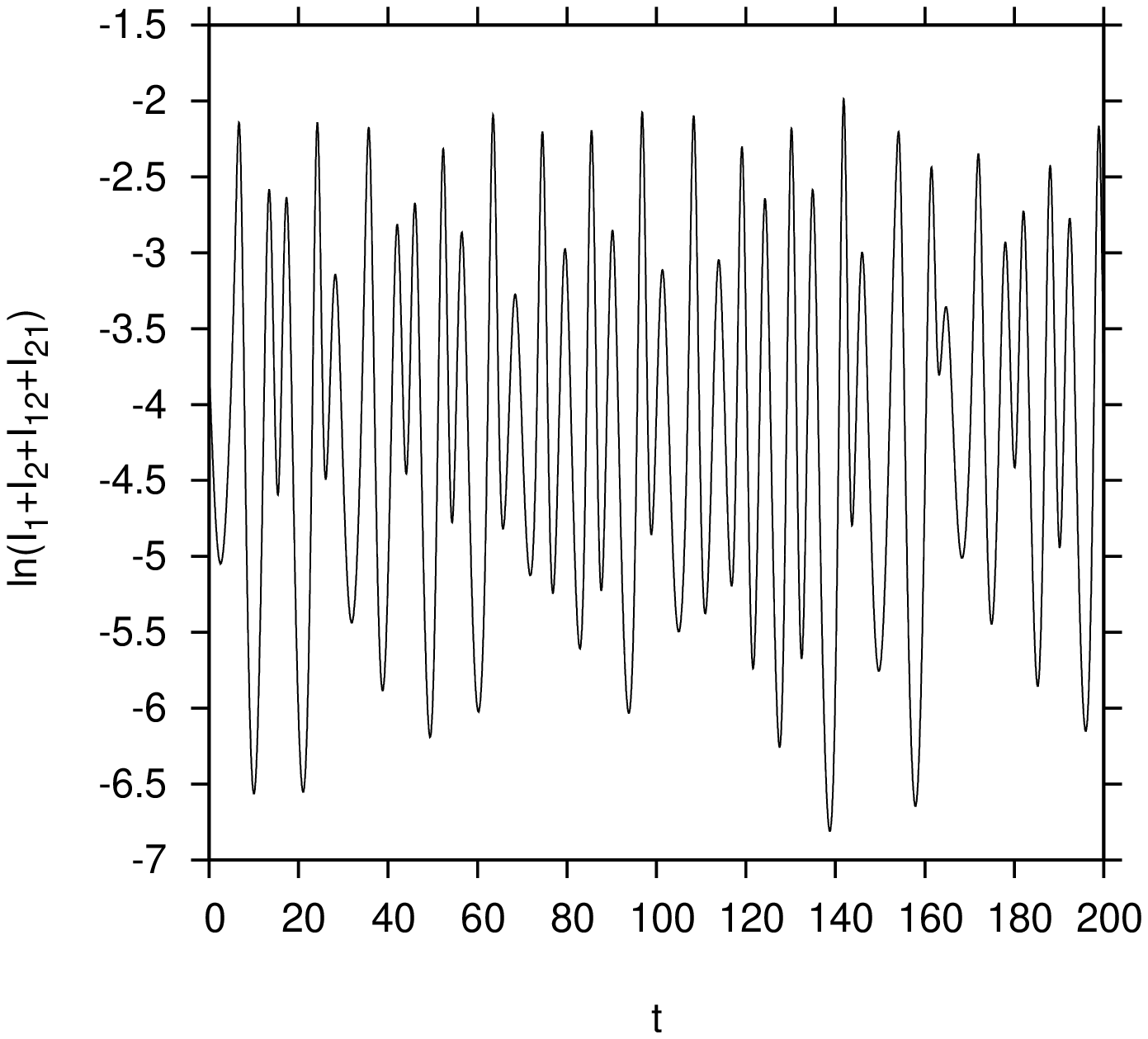}
	{Using the same parameter values above, in a) we plot the time 		  series of the total number of infected
          $I:=I_1+I_2+I_{12}+I_{21} $, and in 
          b) time series of the logarithm of the
          overall infected (ln$(I)$), for the same time
          interval.
	}

For $\phi<1$ the number of infected stays quite away from zero, avoiding the chance of extintion in stochastic systems with reasonable system size.
In Fig. \ref{fig:timeseriesIphi27} b) the logarism of total infected goes as low as $-70$, when $\phi>1$. For $\phi<1$, see Fig.
\ref{fig:timeseriesIphi07} b) the logarism of total infected does not 
pass below $-7$.

This encourages to also look closer to the parameter region of
$\phi < 1 $, the region of ''inverse ADE'', i.e. when dengue patients with
severe disease because of the ADE phenomenon contribute less to the force of infection
due to possible hospitalization, and not more, as previous models 
suggested.


\subsection{State space plots }


In Fig. 
\ref{fig:sirmult1} we plot 500 years of
dynamics to clearly observe the dynamic patterns present.

\epsfigthreevarnew{3cm}{3cm}{3cm}{htb}{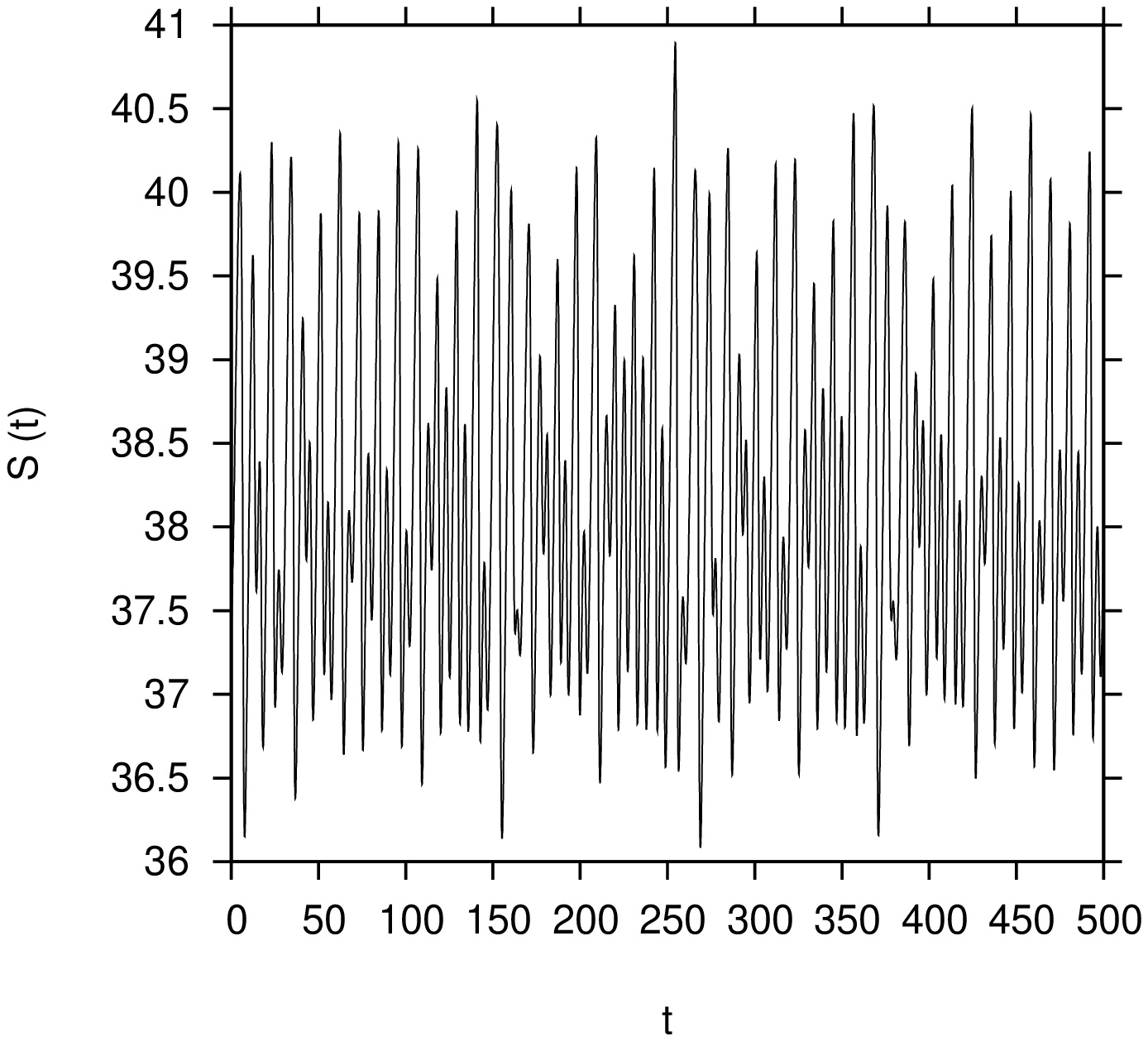}{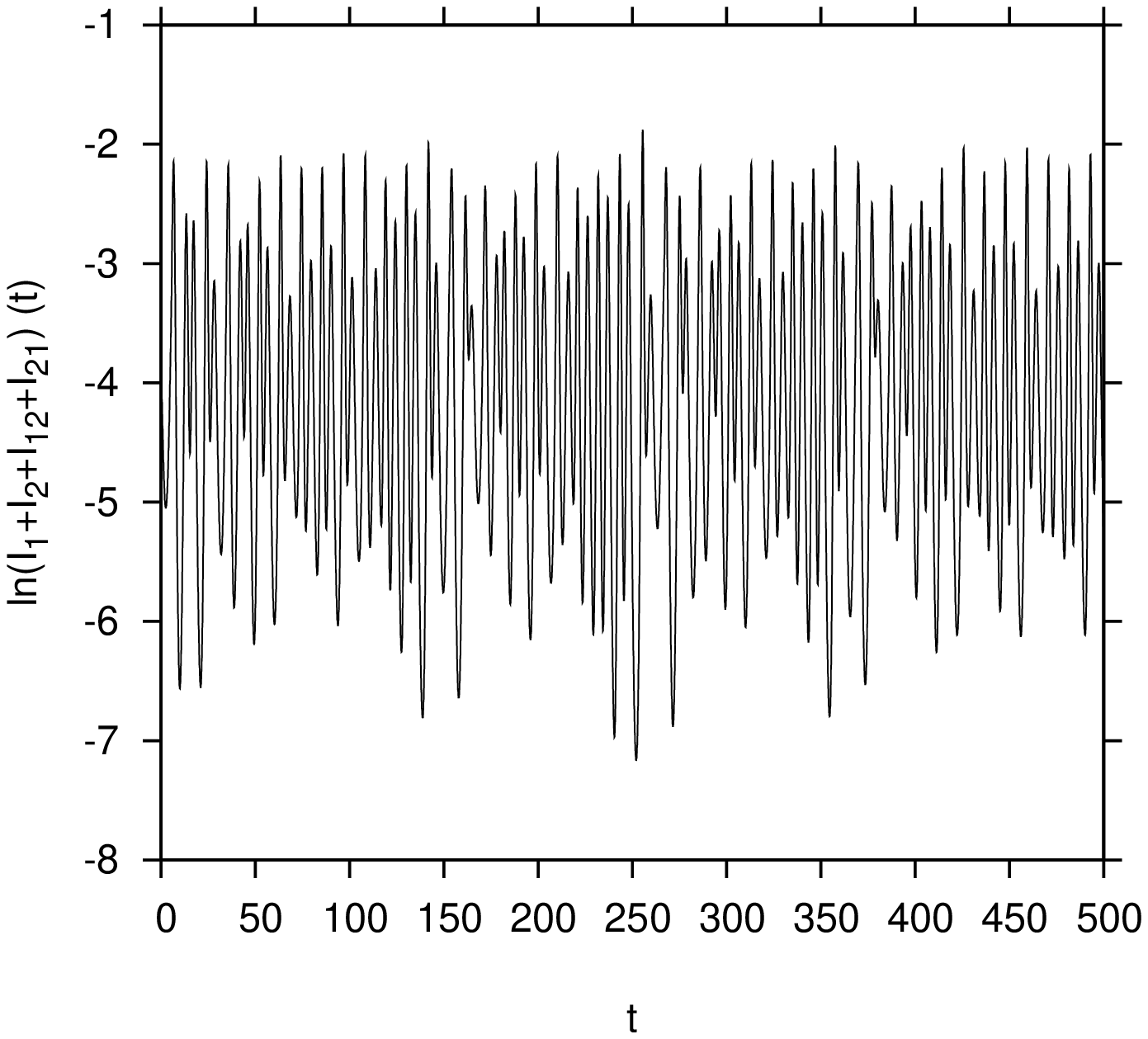}{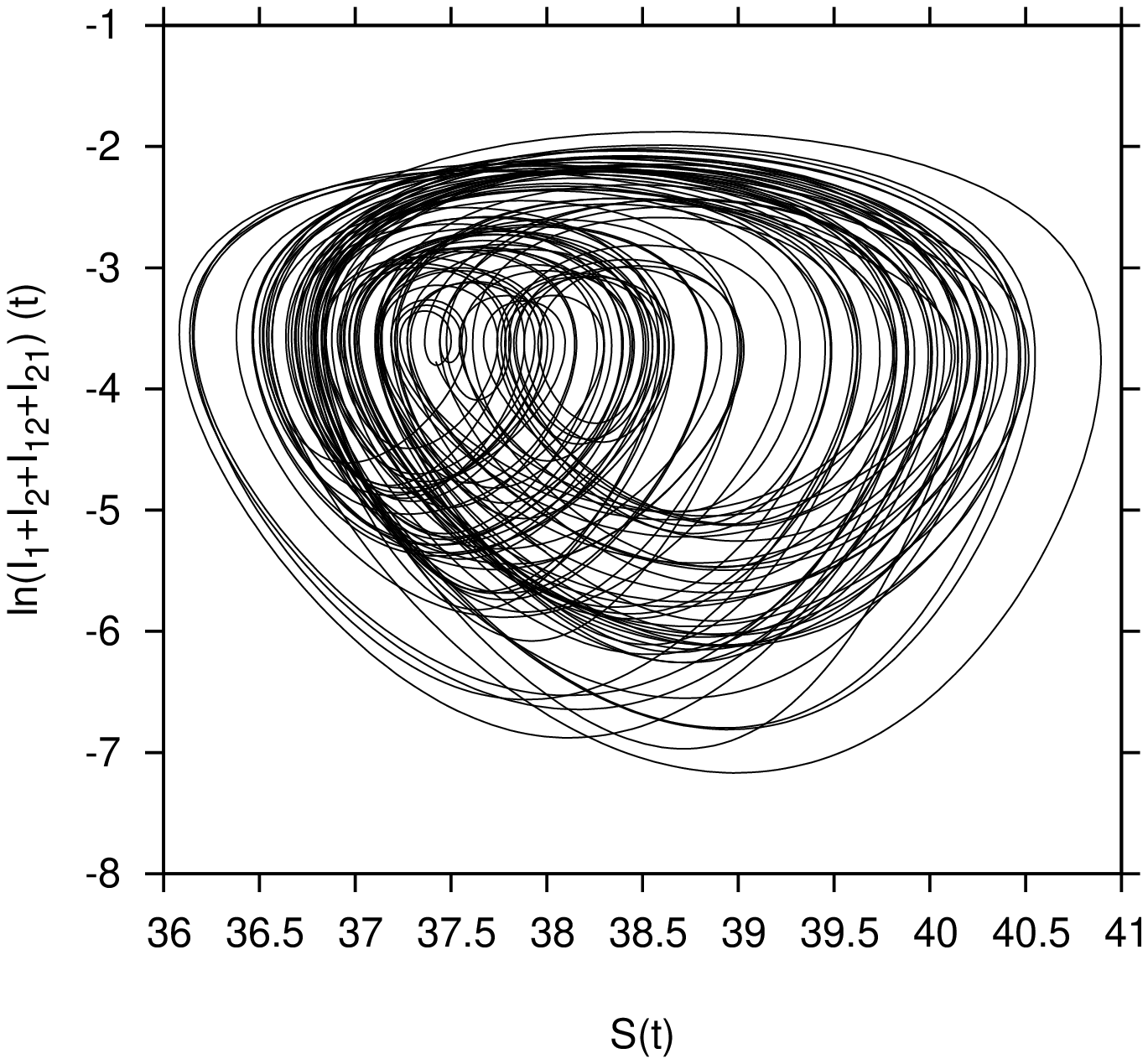}
	{For $\alpha=2$ and $\phi=0.7$, we plot in a) the time series of the 		  susceptibles after discarding a long transient
          (here 2000 years), and in b) the time series of the logarithm 
	  of the
          overall infected, $I:=I_1+I_2+I_{12}+I_{21} $ for the same time
          interval.
         In c) State space plot of the number of susceptibles
          versus logarithm of the
          overall infected, as show from their time series in
          Fig. \ref{fig:sirmult1} a) and b), finding a chaotic attractor.
          Parameters: $\alpha=2 $, $\phi=0.7 $.
	}

We now plot the total number of infected with one strain versus
the total number of infected with the other strain.
In case of synchronicity of the two strains we would expect the
system to stay closer to the main diagonal of the plot, whereas for
anti-synchronicity it would mainly stay near the off-diagonal.
The observed pattern is more supporting anti-synchronicity, but
near-synchronicity can also be observed at times 
(see Fig. \ref{fig:sirmult3neu}).
For quantification of these effects of chaos synchronization or
chaos desynchronization applied to dengue models
see Schwartz {\it et al.} \cite{Schwartz.et.al2005}.

\epsfig{4cm}{htb}{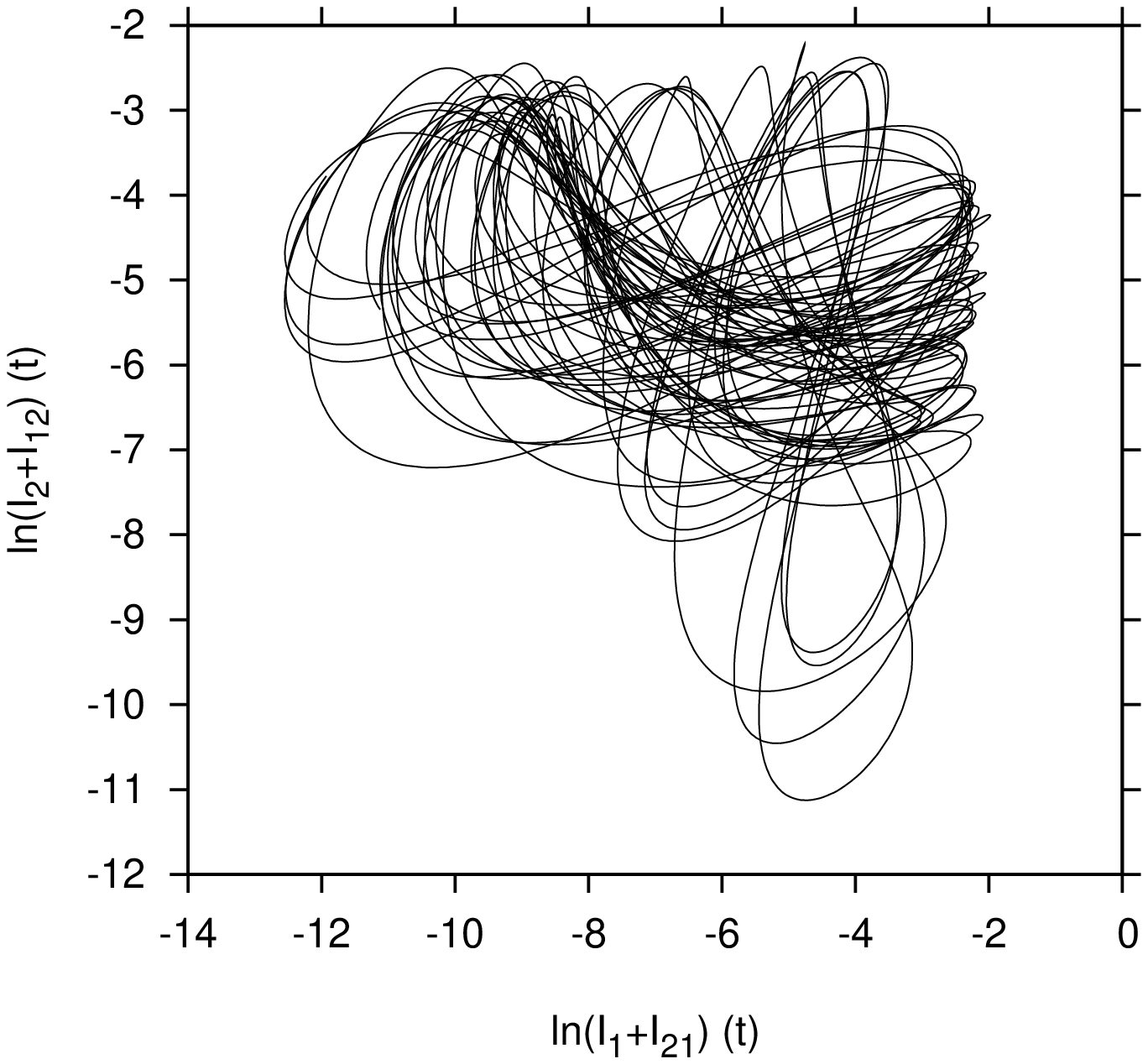}
	{State space plot of the logarithmic number of infected
          with strain 1 versus the ones with strain 2.
          Parameters: $\alpha=2 $, $\phi=0.7 $.
	}

In the further analysis we will investigate the state space
plots in terms of the variables $S $ and the logarithm of
$I:=I_1+I_2+I_{12}+I_{21} $, since
dengue notification data often do not distinguish between 
the circulating strains, hence the total number of infected $I $, 
whereas the susceptible class $S$ is $N $ minus every host who ever has experienced an infection. 

In eventual data analysis the method of delay
coordinates even allows to only work with one time series of
$I$, and analysing $I(t)$, $I(t+\tau ) $ etc. with a time delay
$\tau $ \cite{Packardetal80,FarmerSidorowich}.



\subsection{Bifurcations of limit cycles when changing  {\boldmath $\phi $}}


The state space plots for various values of $\phi $ show bifurcations
from fixed point behaviour to limit cycles, which then bifurcate into
double-limit cycles etc.,

\epsfigfour{3cm}{3cm}{htb}{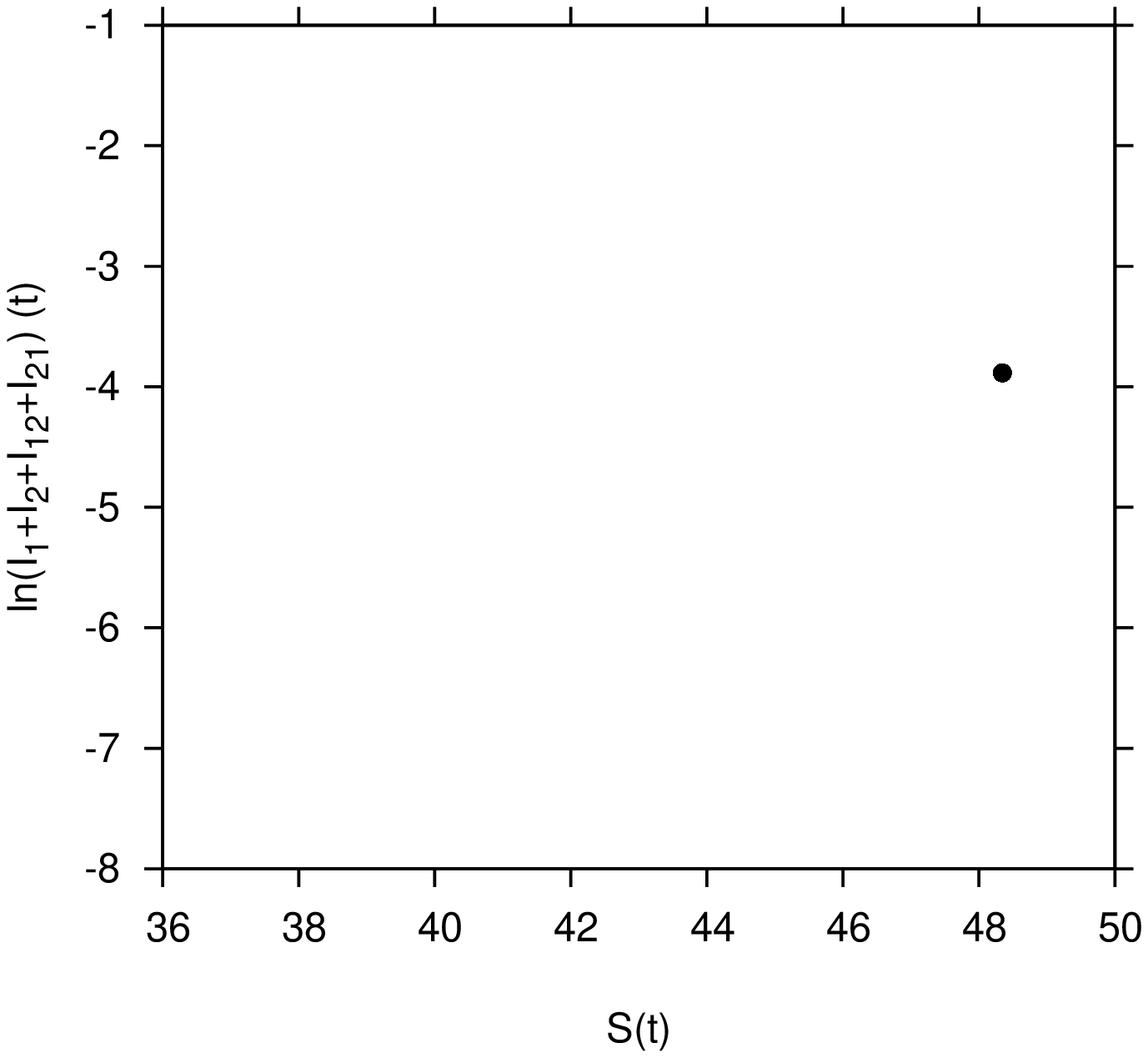}{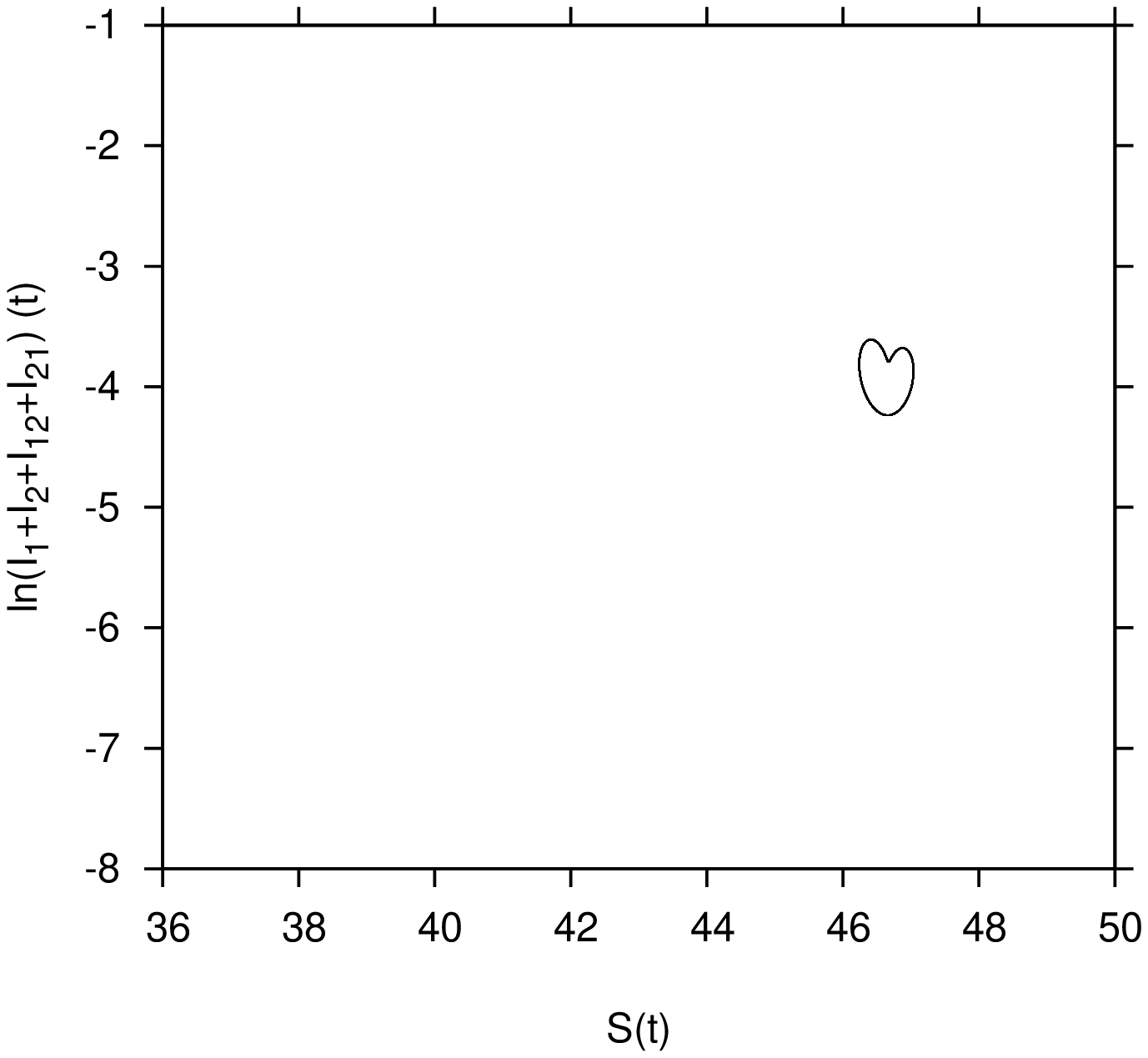}{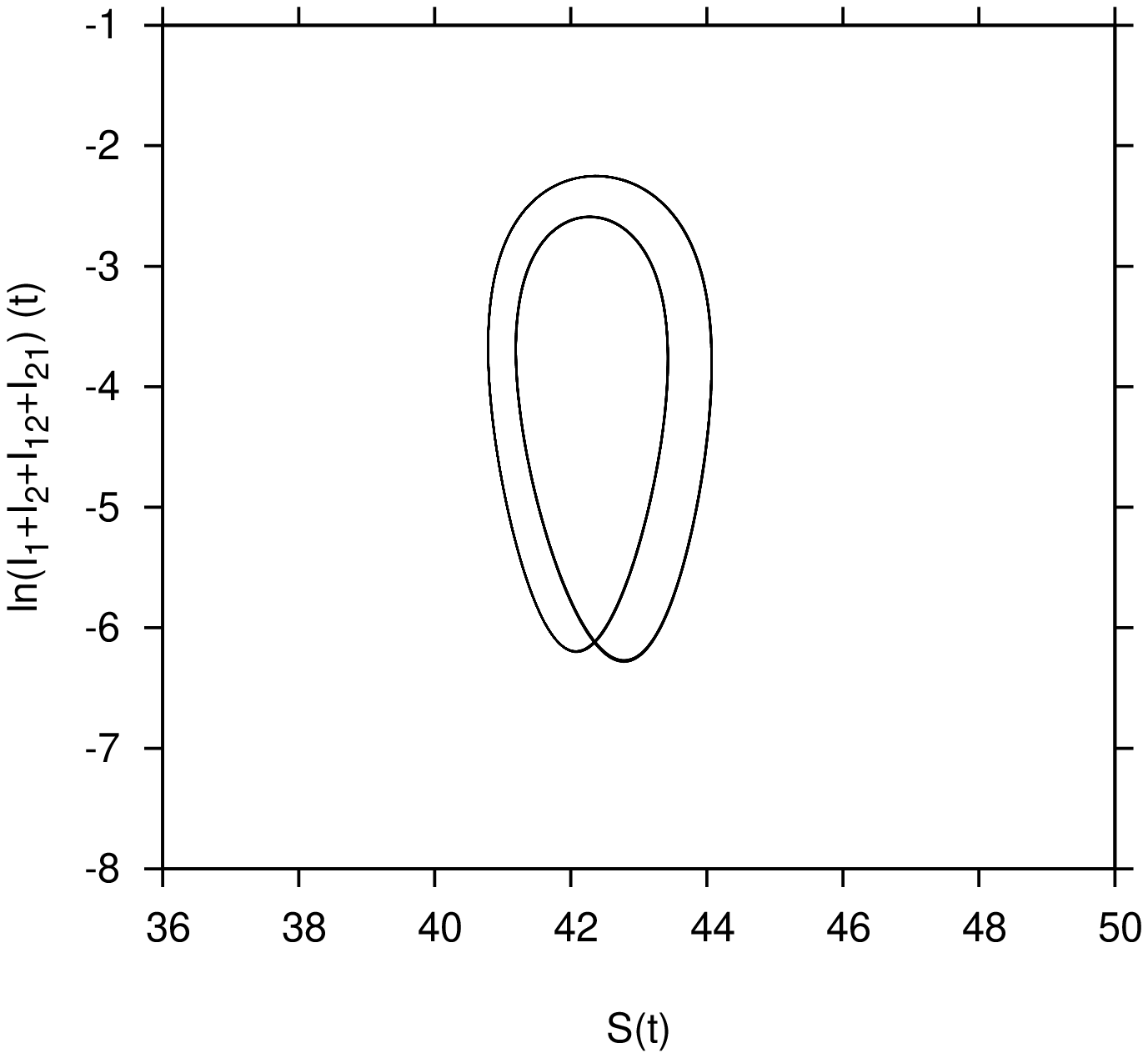}{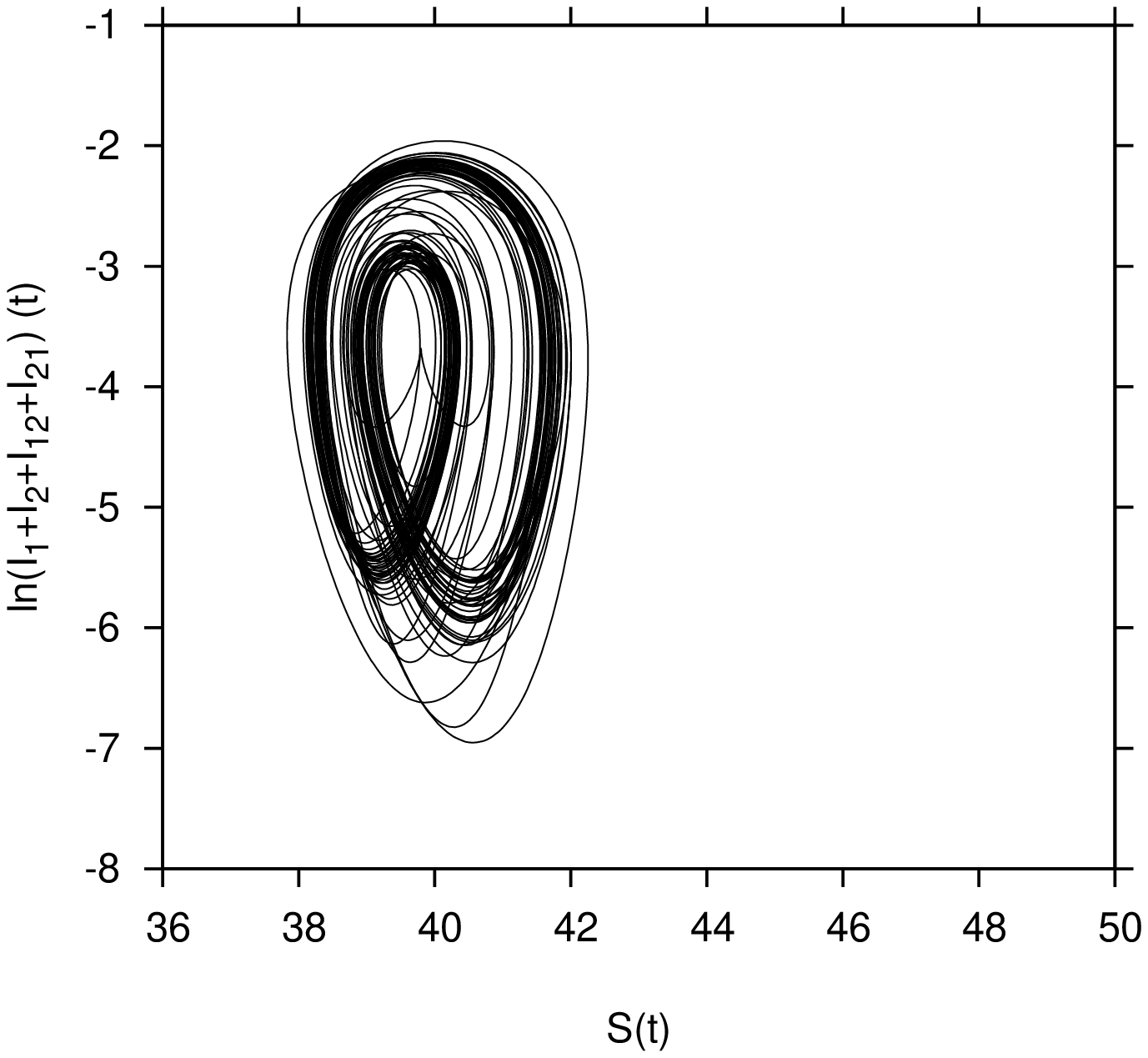}
	{Attractors for various values of $\phi <1 $: a) fixed point for
          $\phi =0.1$, b) limit cycle for $\phi =0.2$, c) bifurcating
          limit cycle for $\phi =0.45$, d) chaotic attractor for
          $\phi =0.6$.
	}

\epsfigfour{3cm}{3cm}{htb}{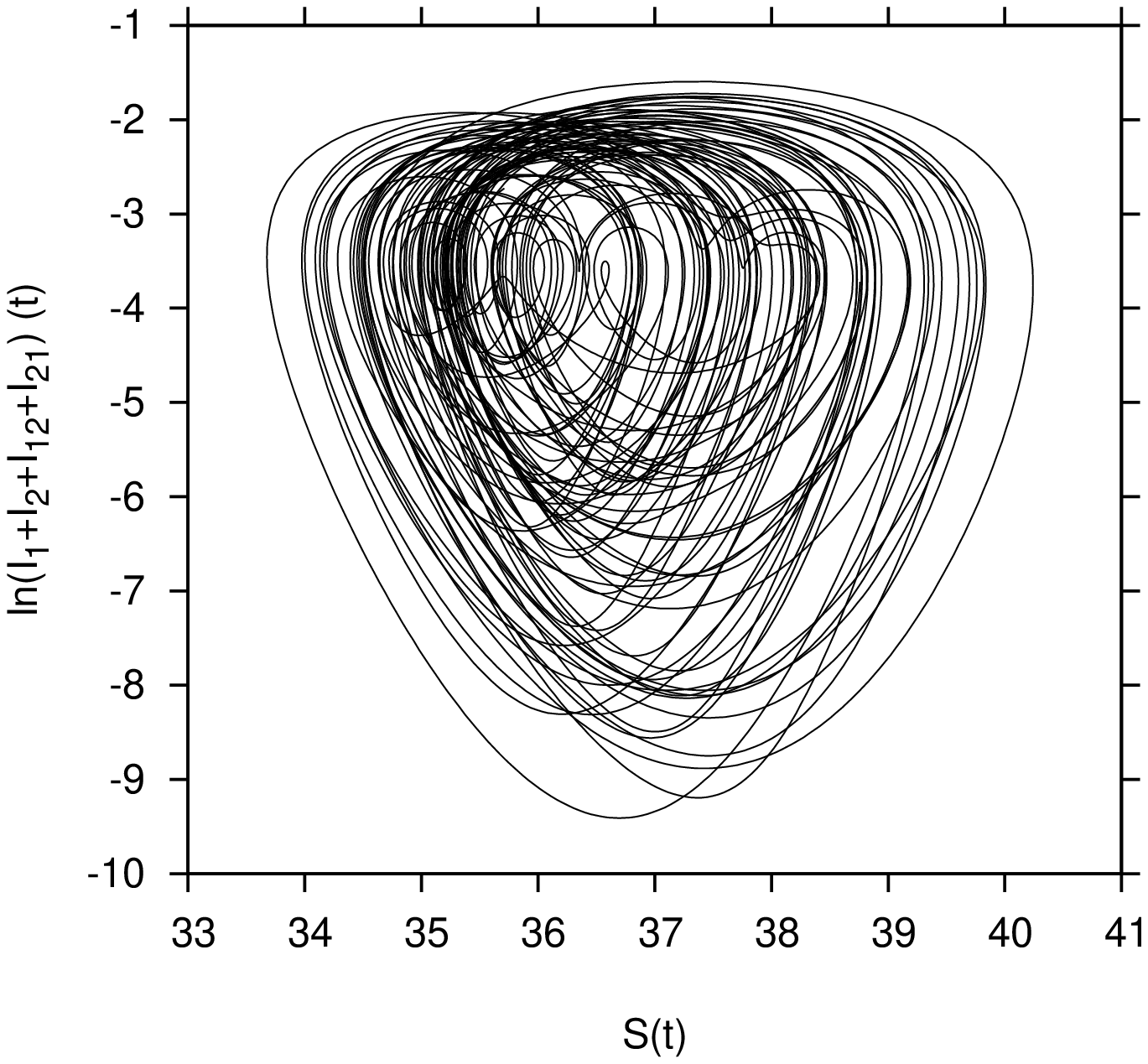}{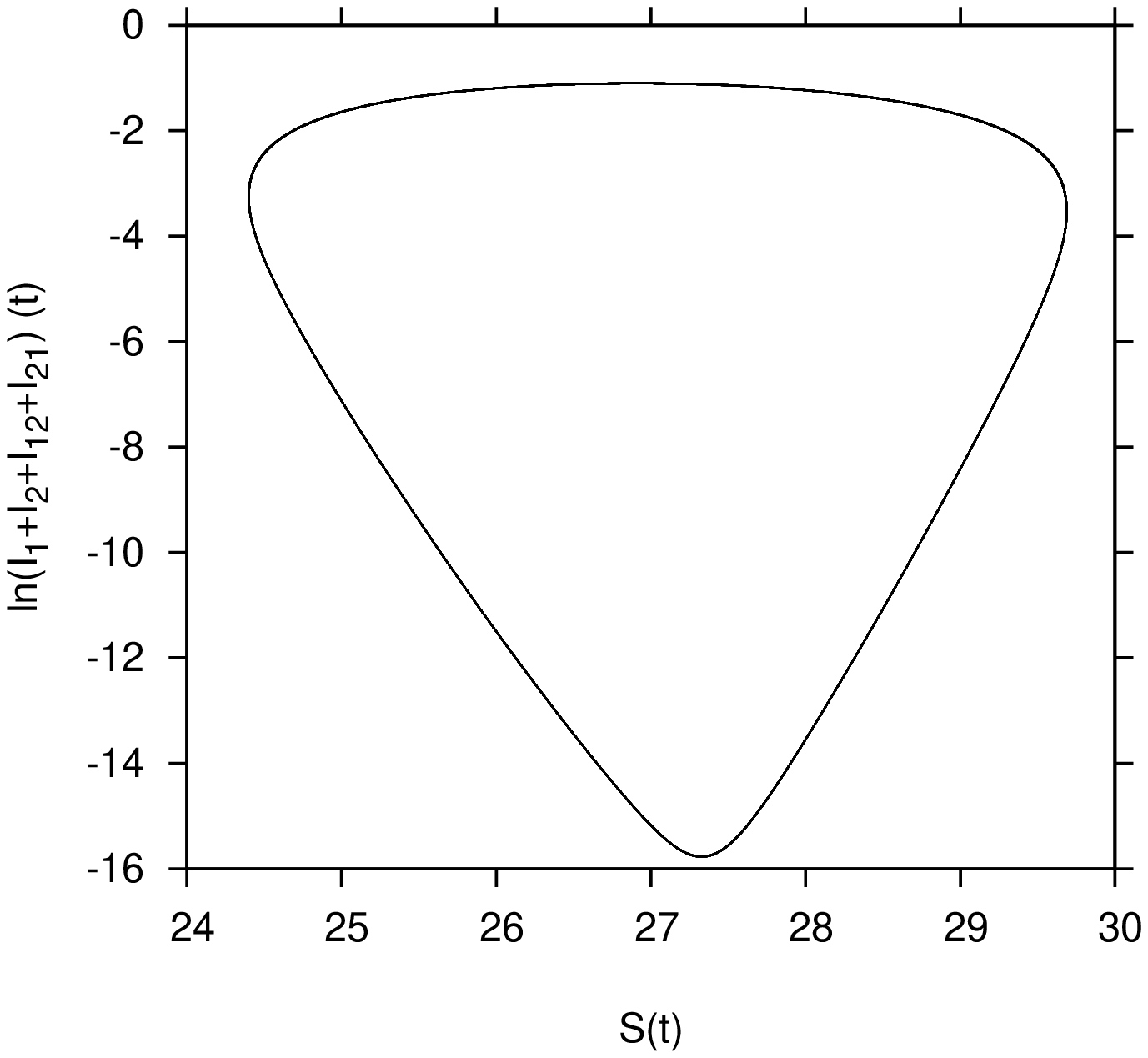}{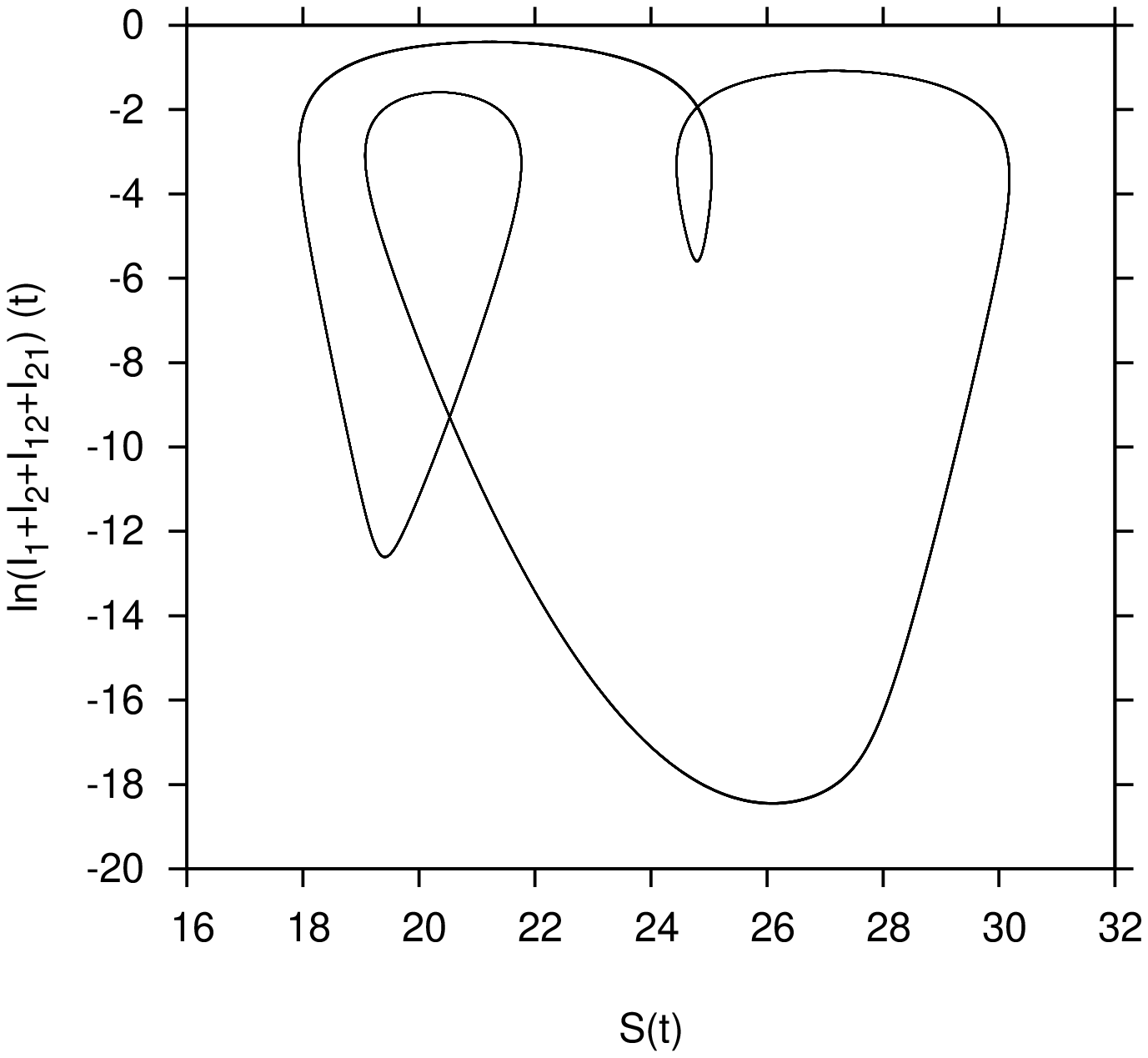}{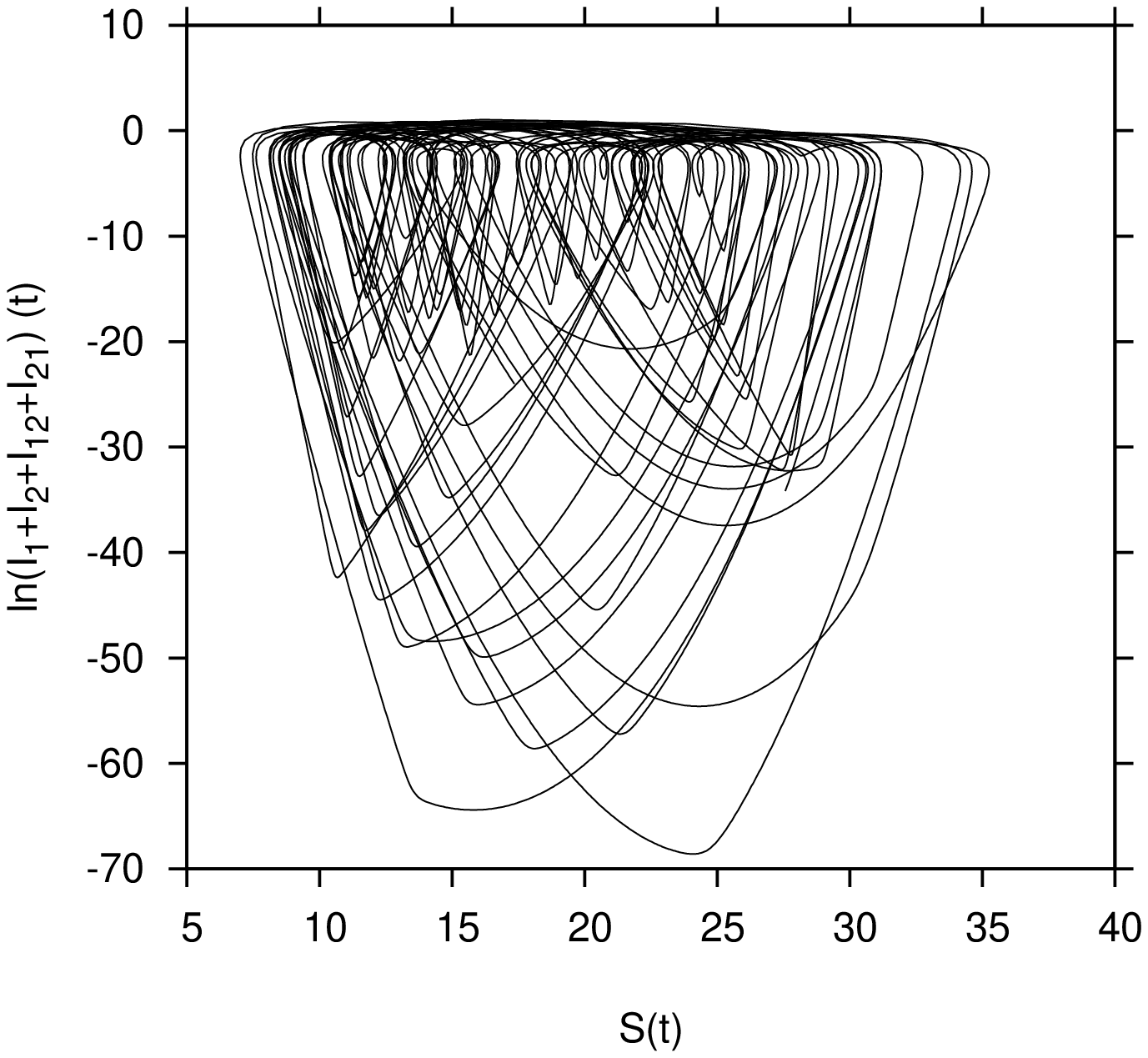}
	{Attractors for various values of $\phi $ around 1 and larger: 
          a) chaotic attractor for
          $\phi =0.8$, b) limit cycle for $\phi =1.5$, c) more complicated
          limit cycle for $\phi =1.9$, d) chaotic attractor for
          $\phi =2.7$.
	}

\clearpage

\noindent
until completely irreguar behaviour, which is the fingerprint of deterministc chaos. See Fig. \ref{fig:fofo2phi01} and Fig. \ref{fig:fofo1phi08}.

Looking for higher values of $\phi $, than just towards the first
chaos window, shows that the chaotic attractor becomes unstable
again, just leaving simple limit cycles as attractors for large
parameter regions beyond of $\phi =1$. Only for much higher
values of $\phi  >> 1$, another chaotic attractor appears, the so-called
ADE chaotic attractor \cite{FergusonN.et.al1999,Schwartz.et.al2005}.



\subsection{Map of maxima of {\boldmath $I $} in state space }


We also investigate maxima maps in order to classify the dynamics for various parameter values. see Fig. \ref{fig:fofo1phi19} and Fig. \ref{fig:sirmult4neu}.

\epsfigtwo{3cm}{3cm}{htb}{fofo1phi19}{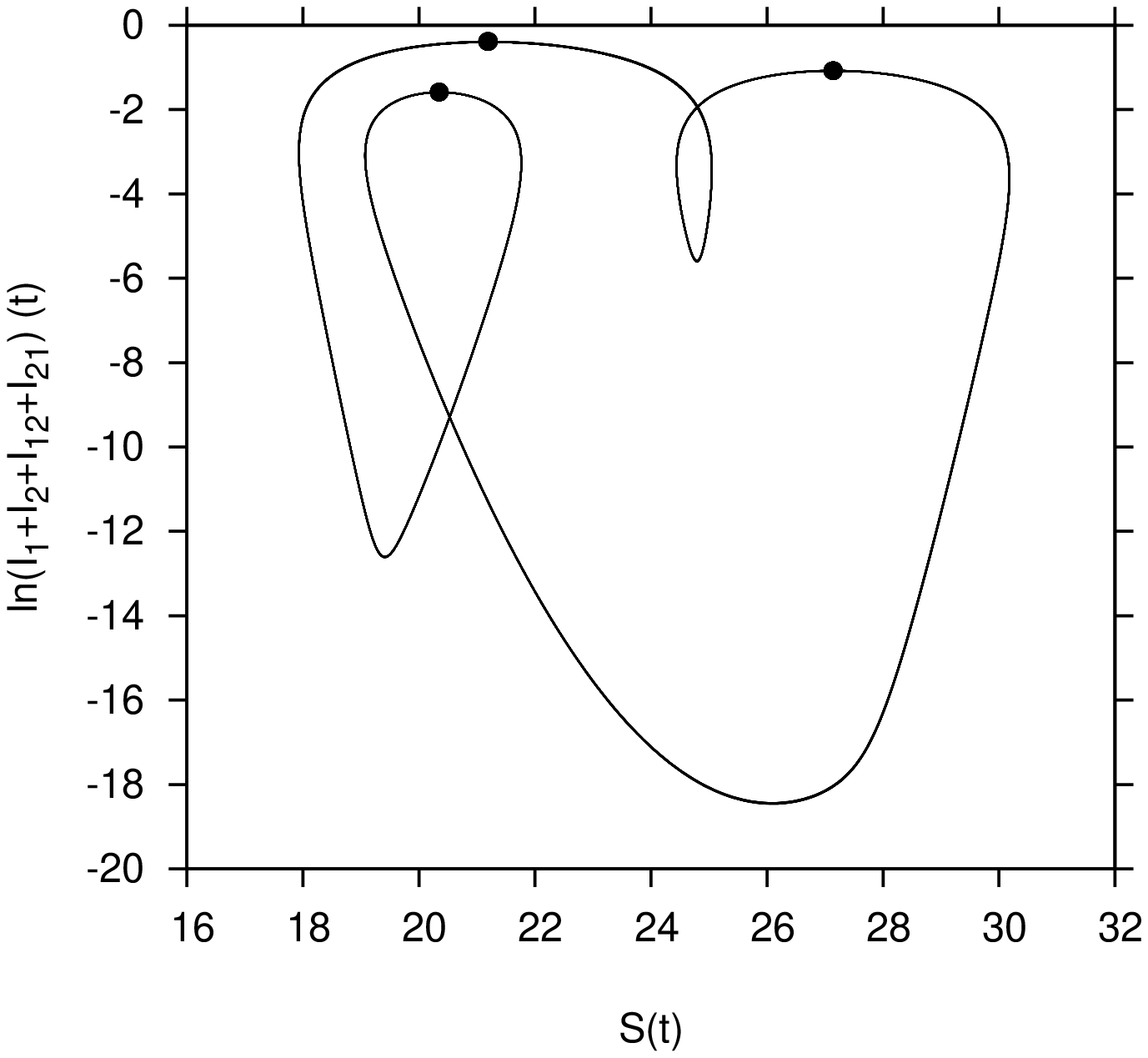}
	{a) Limit cycle for $\phi=1.9 $, b) The dots indicate the
          local maxima calculated for the
          twisted limit cycle of a). The three dots characterize
          the three limit cycle.
          Parameters: $\alpha=2 $, $\phi=1.9 $.
	}


We plot for the time $t_{max} $, at which the total number of
infected $I(t):=I_1+I_2+I_{12}+I_{21} $ has a local maximum, 
the number of infected at that time
$ln(I(t_{max})) $ and for the same time value the susceptibles
$S(t_{max}) $. In this way we obtain a maxima map.

\epsfigthreevarnew{3cm}{3cm}{3cm}{htb}{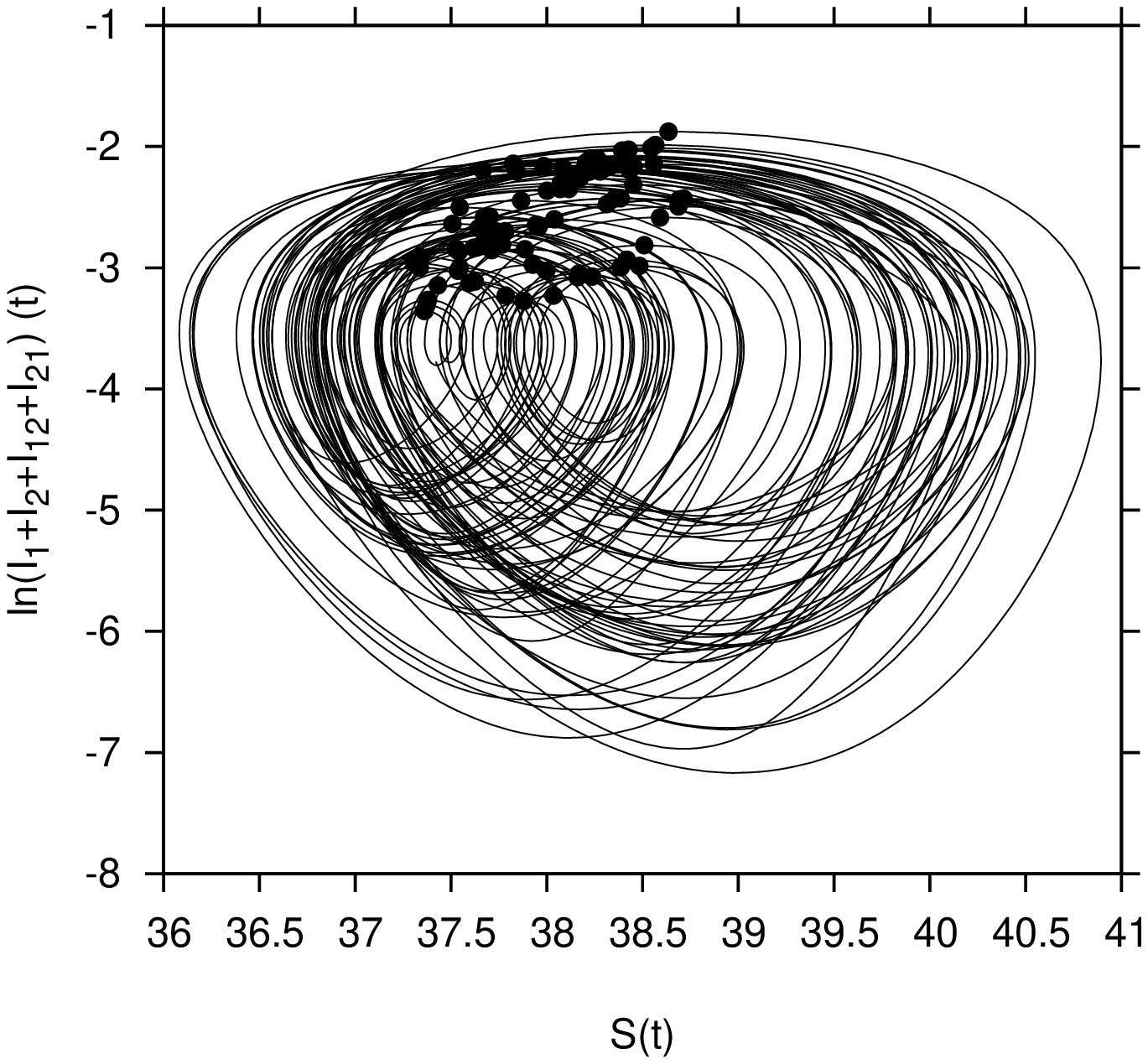}{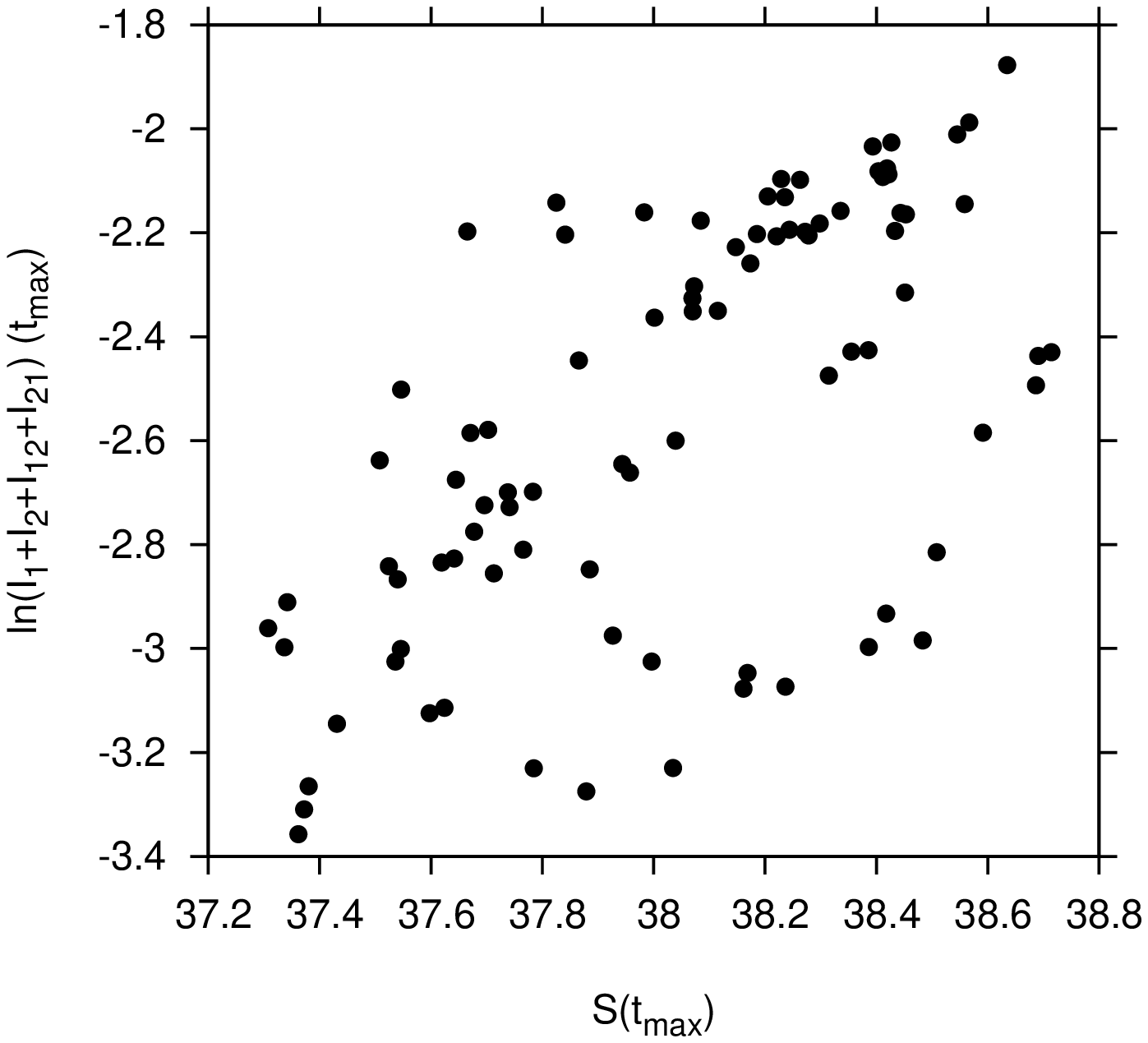}{maximamaplight}
	{In a), the determined local maxima for the chaotuc attractor found 		for $\alpha=2 $ and $\phi=0.7 $ (See Fig. \ref{fig:sirmult1} c)). In 		b), scattered maxima plot for the $500$ years
	of simulation. In
          c) the map of maxima of the overall infected and the respective
          values for the susceptibles for a very long simulation, using
          the same conditions as for 
          Fig \ref{fig:sirmult4neu} b). We observed that even after $200000$ 	       years, the dots never come back to the same point.
          Much more structure than the scatter
          plot appears.
          The fingerprint of the chaotic
          attractor is clearly visible now.
	}

Whereas in Fig. \ref{fig:sirmult4neu} b) we use 500 years of simulations,
observing rather erratical scattering of points of local maxima,
for Fig. \ref{fig:sirmult4neu} c) we use the local maxima of a long
simulation of 200 000 years to observe the clear fingerprint of
deterministically chaotic behaviour.


 \clearpage


\subsection{Bifurcation diagrams }


Ploting the local maxima of $ln(I)$ over the varying parameter $\phi$, 
we obtain a bifurcation diagram,
in which fixed points and simple limit cycles appear as one dot 
per parameter value, whereas double-limit cycles appear as two
dots, more complicated limit cyles as more dots, and chaotic attractors
as continuously distributed dots for a single $\phi$ value 
(see Fig. \ref{fig:sirmult6v4light}).


\epsfig{4cm}{htb}{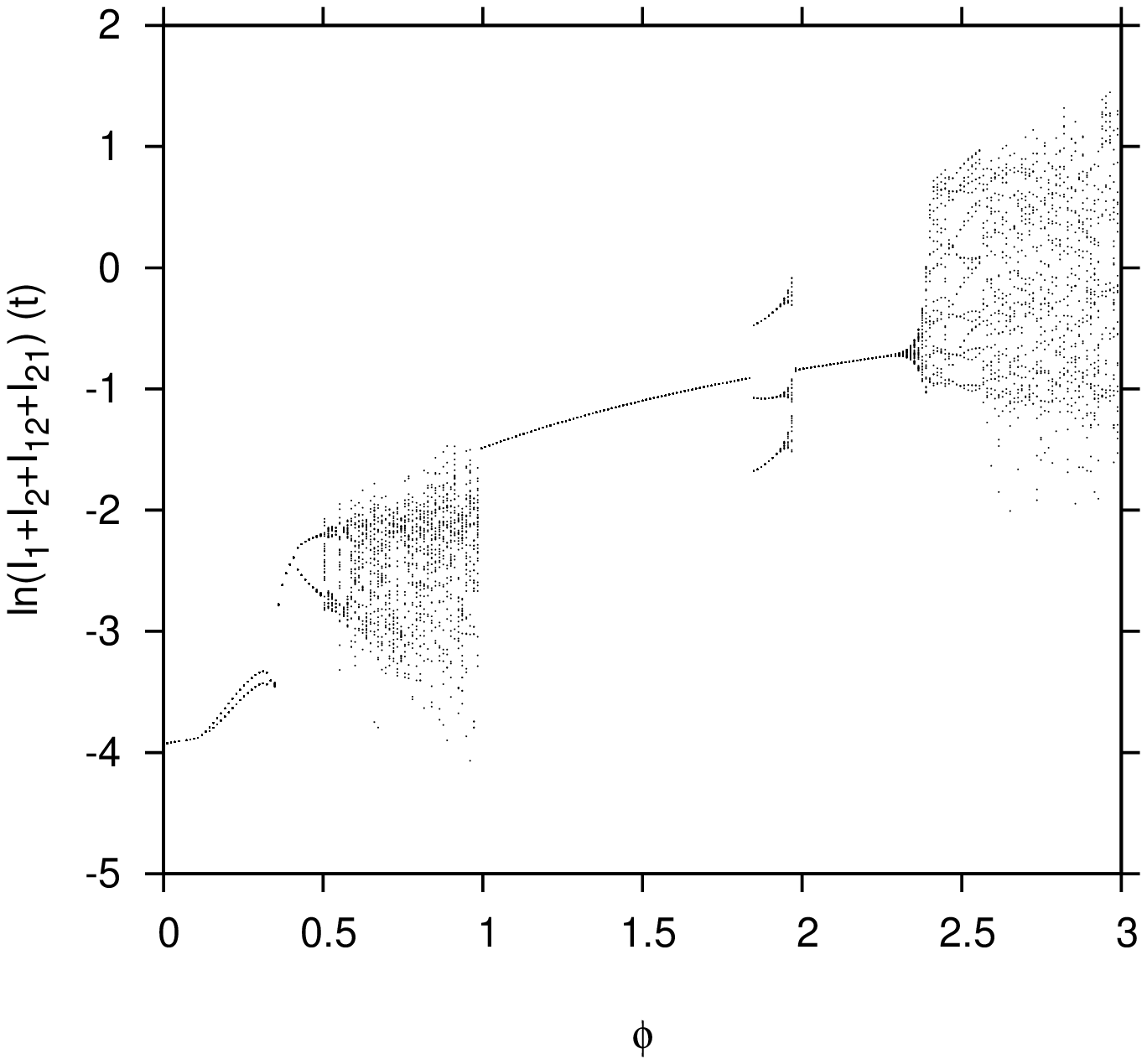}
	{Bifurcation diagram for the maxima of the overall infected
          with changing parameter $\phi $ and $\alpha =2 $.
          We observed a chaotic window for $\phi<1$, 
          region of "inverse ADE", where this dynamical behaviour 
          has never been described before, and another 
          one for $\phi>1$, called by ADE chaotic window, 
          found in previous publications 
          \cite{FergusonN.et.al1999,Schwartz.et.al2005}.
	}


To be sure that this unexpeted behaviour for $\phi<1$ not just appears because of this specific $\alpha$ value, we look at the robustness of the findings with varying
the temporary cross-immunty parameter values. 

For $\alpha = 1$ e.g, both chaotic windows appears, and surprisingly in the region of "inverse ADE" this window is even larger (see Fig. 
\ref{fig:bifurcationalpha1light}).

\epsfig{4cm}{htb}{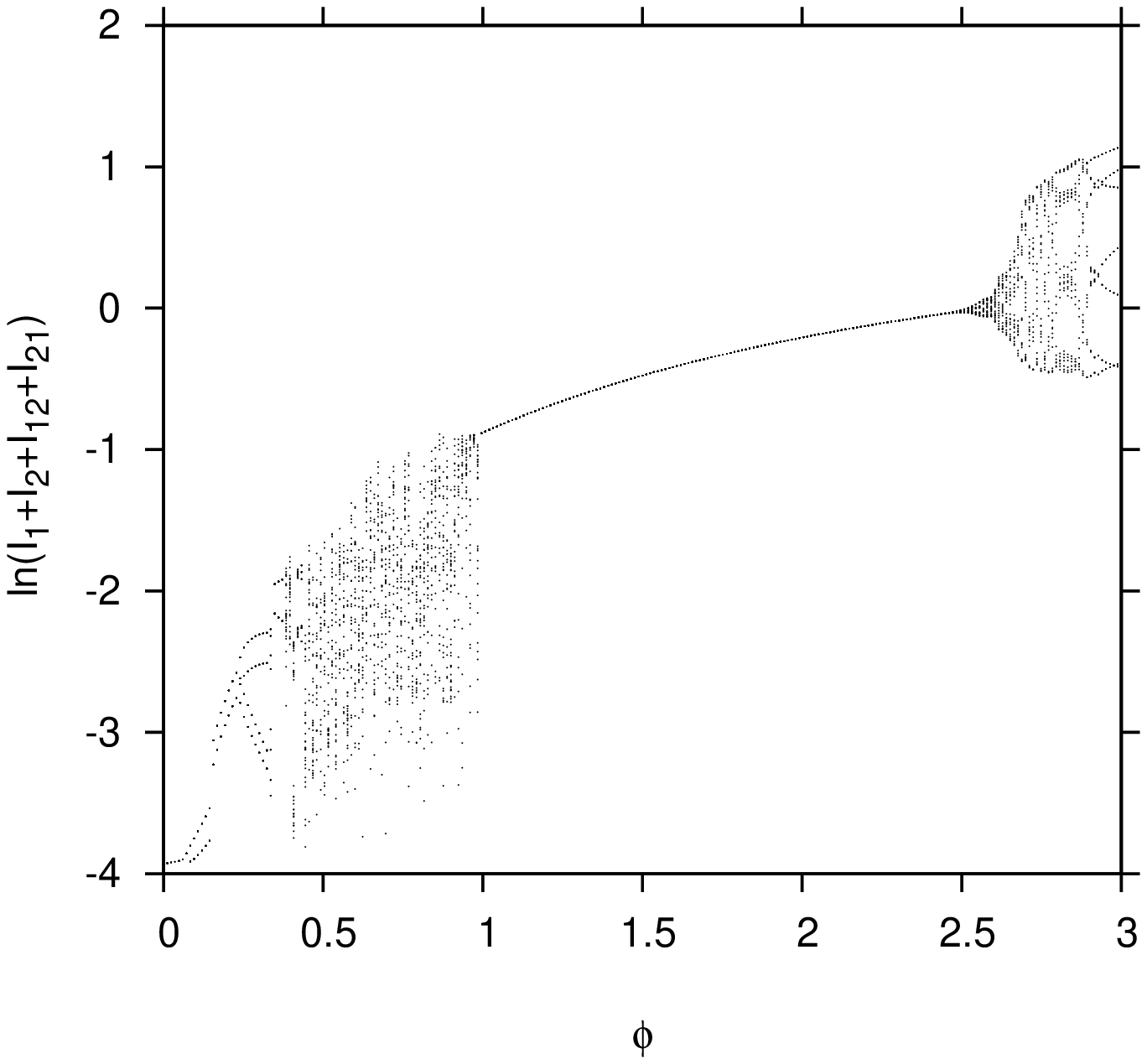}
	{Bifurcation diagram for the maxima of the overall infected
          with changing parameter $\phi $. Now $\alpha =1 $. The bifurcation
          diagram appears to be quite robust against changes of
          parameters around the region under investigation.
	}

Only for very large values of $\alpha $, where temporary cross-immunity
becomes unimportant due to the low resident times in the classes $R_1 $
and $R_2 $, the chaos window for $\phi <1  $ disappears, and then ADE as increasing infectivity on a secondary infection condition
seems to be the only mechanism to observe deterministic chaos (see Fig.
 \ref{fig:bifurcationalpha10light}).

\epsfigtwo{4cm}{4cm}{htb}{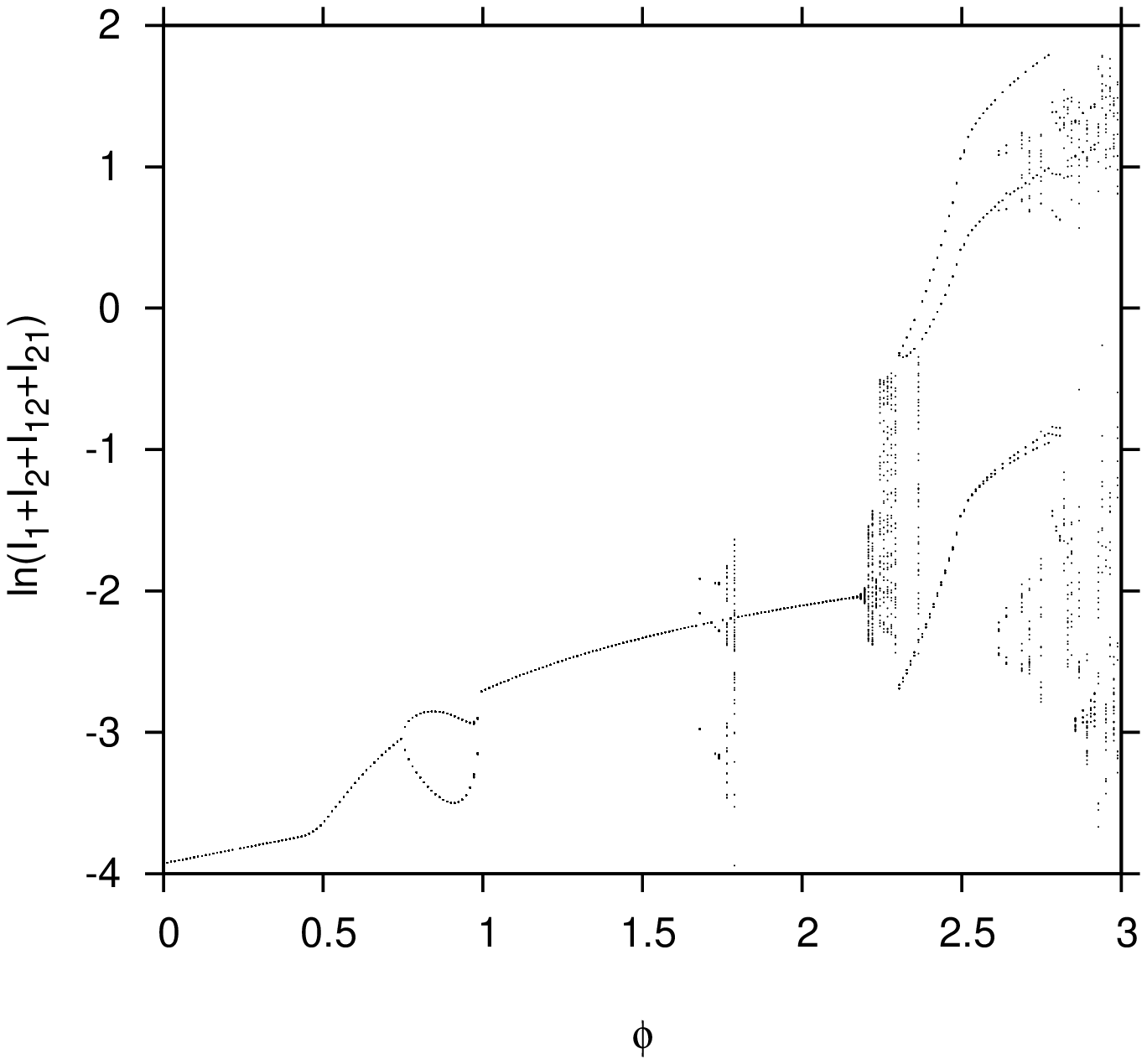}{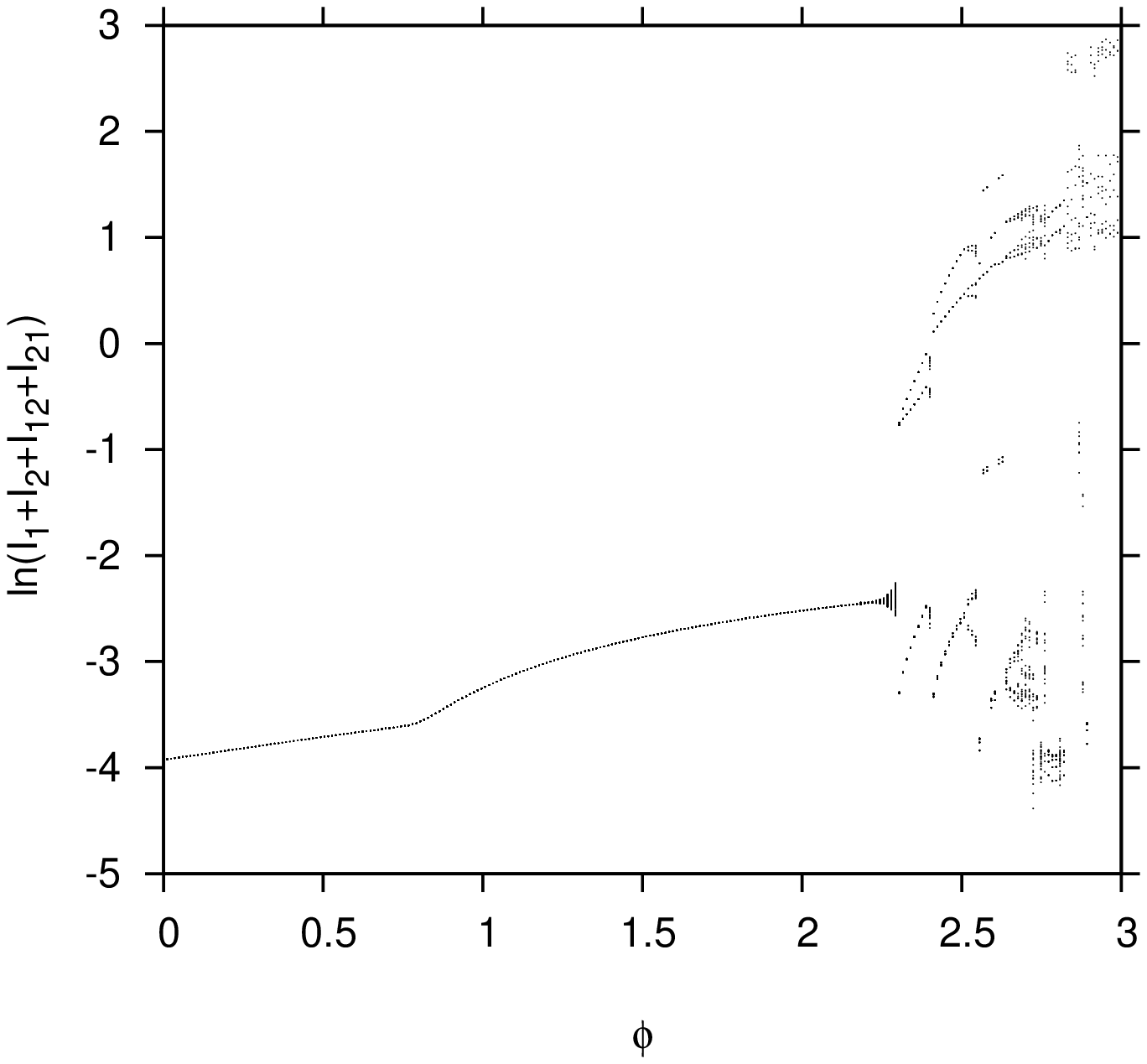}
	{Bifurcation diagram for the maxima of the overall infected
          with changing parameter $\phi $. Now for a very large
          value of $\alpha$ we get close to the models found in the literature, where the temporary cross-immunity is not considered. a) For $\alpha =10 $,
          The chaos window for $\phi <1 $ disappears completely as
          the temporary cross-immunity becomes shorter or unimportant.
          b) For $\alpha =20 $ it becomes clear that there is no
          other dynamics in the region for $\phi <1 $, than fixed
          points or simple limit cycles.
	}

\epsfig{5.0cm}{htb}{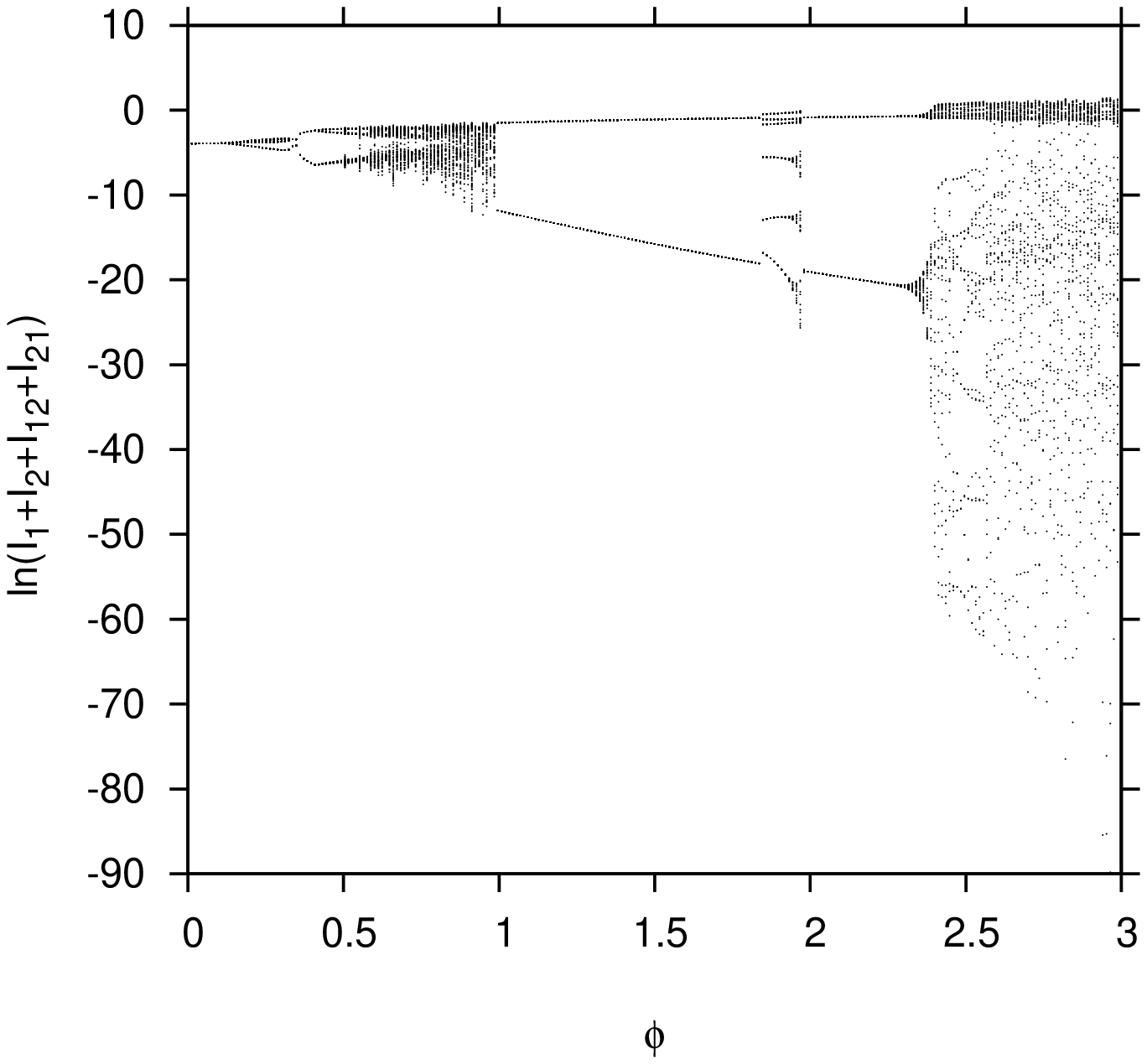}
	{Bifurcation diagram for the extrema of the overall infected
          with changing parameter $\phi $. $\alpha =2 $.
          For the chaotic window for $\phi >1 $ the minimal values go
          to very low numbers of infected.
	}

Finally, plotting not just the local maxima but also the local minima of 
$ln(I)$ in a bifurcation plot,
as done in Fig. \ref{fig:denguebifextremalight}, we observed again for
$\phi >1 $ very low troughs of infected, whereas in the chaotic region
for $\phi <1 $ the overall number of infected stays always sustainably
high.


\section{Summary }


Our analysis shows that also for "inverse ADE", i.e. $\phi <1  $, 
deterministic chaos can be observed, when taking the 
temporary cross-immunity reported in the medical literature 
into account.
  
This indicates that deterministic chaos is much more important
in multi-strain models than previously thought. We could show this
in a very basic model with only two strains and one reinfection
possible, not needing the ADE mechanism, but rather stating that
upon second infection hosts spread a disease less likely, since it might
be more harmfull. This mechanism could be present in other 
diseases than dengue fever, hence of much wider interest.


\section{Acknowledgments}  


We would like to thank Gabriela Gomes, Lisbon, for valuable discussions 
on multi-strain dynamics, Francisco Lemos 
and S\^onia Diniz, Belo Horizonte, for valuable 
information about dengue epidemiology.




\begin{thebibliography}{99}




\bibitem{RuelleTakens}
	Ruelle, D., \& Takens, F. (1971) On the Nature of Turbulence, 
	{\it Commun. Math. Phys.} {\bf 20}, 167--192. See also  
	{\it Commun. Math. Phys.} {\bf 23} (1971), 343--344.

\vspace*{-0.2cm}

\bibitem{Ott}
	Ott, E. (1993) {\it Chaos in Dynamical Systems}
	(Cambridge University Press, Cambridge).


\vspace*{-0.2cm}


\bibitem{Ruelle89}
	Ruelle, D. (1989) {\it Chaotic Evolution and Strange Attractors}
	(Cambridge University Press, Cambridge).

\vspace*{-0.2cm}

\bibitem{Halstead1982}
	Halstead, S. B. (1982)
	Immune enhancement of viral infection. Progress in Allergy.
 	{\it Progress in Allergy}. {\bf 31}, 301--64.

\vspace*{-0.2cm}

\bibitem{Halstead2003}
	Halstead S. B. (2003)
	Neutralization and antibody-dependent enhancement of dengue viruses.
 	{\it Advances in Virus Research}. {\bf 60}, 421--67.

\vspace*{-0.2cm}


\bibitem{FergusonN.et.al1999}
	Ferguson, N. et al. (1999)
	The effect of antibody-dependent enhancement 
        on the transmission dynamics and 
        persistence of multiple-strain pathogens.
	{\it Proc. Natl. Acad. Sci}. {\bf 96}, 790--94.

\vspace*{-0.2cm}

\bibitem{Schwartz.et.al2005}
	Schwartz, I.B., Shaw, L.B., Cummings, D.A.T., Billings, L.,
        McCrary, M., Burke, D.S. (2005)
	Chaotic desynchronization of multi-strain diseases.
	{\it Physical Review}. {\bf E 72}, 066201--6.

\vspace*{-0.2cm}

\bibitem{HelenRohani2006}
        Wearing, H.J.,\&  Rohani, P. (2006)
        Ecological and immunological determinants of dengue epidemics, 
        {\it PNAS } {\bf 103}, 11802--11807.

\vspace*{-0.2cm}

\bibitem{WHO2002}
	Word Health Organization. (2002) 
	{\it Dengue and Dengue Hemorrhagic Fever}. 
        (World Health Org., Geneva, Fact Sheet {\bf 117}).

\vspace*{-0.2cm}

\bibitem{Favier.et.al2005}
	Favier, C. et al. (2005) 
	Influence of spatial heterogeneity on an emerging   
        infectious disease: the case of dengue epidemics.
 	{\it Proc. Biol. Sci}. {\bf 272 }, 1171--7.

\vspace*{-0.2cm}

\bibitem{Monath1994}
        Monath T. P. (1994) Dengue: The risk to developed 
        and developing countries. 
        {\it Proc. Natl. Acad. Sci. U.S.A.} {\bf 91}, 2395--2400.

\vspace*{-0.2cm}

%

\bibitem{Rigau-Perez.et.al1998}
	Rigau-P\'erez, J. G. et tal. (1998)
	Dengue and dengue haemorrhagic fever.
 	{\it The Lancet}. {\bf 352}, 971--77.

\vspace*{-0.2cm}

\bibitem{Rosen.et.al1983}
	Rosen, L. et tal. (1983)
	Transovarial transmission of dengue viruses by mosquitoes: 
        A. albopictus and A. aegypti.
 	{\it Am. J. Trop. Med. Hyg}. {\bf 32}, 1108--19.

\vspace*{-0.2cm}

\bibitem{MortimerR}
	Mortimer, R. Aedes aegypti and Dengue fever.
	{\bf Access at:} \\
	http://www.microscopy-uk.org.uk/mag/art98/aedrol.html

\vspace*{-0.2cm}


\bibitem{DerouichBoutayebTwizell2003}
	Derouich, M., Boutayeb, A., \& Twizel, E. H. (2003) 
	A model of dengue fever.
 	{\it BioMedical Engineering OnLine}. 
        {\bf Access at:} \\
        http://www.biomedical-engineering-online.com/content/2/1/4

\vspace*{-0.2cm}

\bibitem{Huan-Yao.et.al2001}
	Lei, Huan-Yao. et al. (2001)
	Immunopathogenesis of Dengue Virus Infection.
 	{\it Biomedical Science}. {\bf 8}, 337--88.

\vspace*{-0.2cm}

\bibitem{SES}
        Francisco Lemos, pers. comm., zoonosis control coordinator of 		Secretaria de Estado de Sa\'ude de Minas Gerais, Brazil, and S/ônia 		Diniz, pers. comm., responsible of the virology and rackittsioses 	  service from the Fundação Ezequiel Dias, Minas Gerais, Brazil.


\vspace*{-0.2cm}

\bibitem{StephensonJR2005}
	Stephenson, J. R. (2005)
	Understanding dengue pathogenesis: implications for vaccine design. 
  	{\it Bull. World Health Organ}. {\bf 83}, 308--14.

\vspace*{-0.2cm}

\bibitem{FischerHalstead1970}
	Fischer, D. B., \& Halstead, S. B. (1970)
	Observations related to pathogenesis of dengue hemorrhagic fever. 
        V. Examination of age specific sequential infection rates 
        using a mathematical model.  	  
        {\it J. Biol. Med}. {\bf 42}, 329--49.

\vspace*{-0.2cm}


\bibitem{Packardetal80}
	Packard, N.H., Crutchfield, J.P., Farmer, J.D. \& Shaw, R.S. (1980)
	Geometry from a Time Series,
	{\it Phys. Rev. Lett.} {\bf 45}, 712--716 .

\vspace*{-0.2cm}

\bibitem{Takens81}
	Takens, F. (1981) Detecting strange attractors in turbulence,
	in: {\it Dynamical Systems and Turbulence, Warwick 1980} eds.
	D. Rand, L.S. Young
	{\it Lecture Notes in Mathematics} {\bf 898} (Springer, Berlin)
	366 ff.

\vspace*{-0.2cm}

\bibitem{vanKampen} 
	van Kampen, N. G. (1992). {\it Stochastic Processes in Physics 
	and Chemistry}
	(North-Holland, Amsterdam).

\vspace*{-0.2cm}

\bibitem{AlonsoMcKanePascual} 
        Alonso, D.,  McKane, A., \&  Pascual, M. (2006)
        Stochastic Amplification in Epidemics,
        {\it Journal of the Royal Society Interface}, 
        FirstCite Early Online Publishing (not yet in paper): 
        DOI: 10.1098/ rsif.2006.0192. 

\vspace*{-0.2cm}

\bibitem{McKaneNewman}
        McKane, A. J. \& Newman, T. J. (2005)
        Predator-prey cycls from resonant amplification of demographic 
        stochasticity, 
        {\it Phys. Rev. Lett.} {\bf 94}, 218102--7.

\vspace*{-0.2cm}

\bibitem{FarmerSidorowich} 
	Farmer, J.D. \& Sidorowich, J.J. (1987) 
	Predicting chaotic time series,
        {\it Phys. Rev. Lett. } {\bf 95}, 845--848. 





\end{thebibliography}
\end{document}